\documentclass[12pt]{article}
\usepackage{booktabs}

\usepackage{threeparttable}
\usepackage[colorlinks,bookmarksopen,bookmarksnumbered,citecolor=blue,urlcolor=blue,linkcolor=blue]{hyperref}
\usepackage{float}
\usepackage{verbatim} 
\usepackage{amsmath}
\usepackage{enumerate}

\usepackage{graphicx}
\graphicspath{{Pictures/}}
\allowdisplaybreaks[4]

\def\spacingset#1{\renewcommand{\baselinestretch}%
	{#1}\small\normalsize} \spacingset{1}

\usepackage{amsthm,amsmath,enumerate,amsbsy,amsfonts,amssymb,mathabx,amscd,graphicx,algorithm, indentfirst}
\usepackage{natbib}
\usepackage{geometry}
\usepackage{authblk, caption, subcaption}
\usepackage{mathrsfs}
\usepackage{epsfig}

\newtheorem{proposition}{Proposition}

\newtheorem{example}{Example}
\newtheorem{condition}{Condition}

\usepackage{multirow}

\newcommand{\argmin}{\mathop{\mathrm{argmin}}}
\newcommand{\argmax}{\mathop{\mathrm{argmax}}}

\makeatletter
\DeclareRobustCommand\widecheck[1]{{\mathpalette\@widecheck{#1}}}
\def\@widecheck#1#2{%
	\setbox\z@\hbox{\m@th$#1#2$}%
	\setbox\tw@\hbox{\m@th$#1%
		\widehat{%
			\vrule\@width\z@\@height\ht\z@
			\vrule\@height\z@\@width\wd\z@}$}%
	\dp\tw@-\ht\z@
	\@tempdima\ht\z@ \advance\@tempdima2\ht\tw@ \divide\@tempdima\thr@@
	\setbox\tw@\hbox{%
		\raise\@tempdima\hbox{\scalebox{1}[-1]{\lower\@tempdima\box
				\tw@}}}%
	{\ooalign{\box\tw@ \cr \box\z@}}}
\makeatother
\RequirePackage[colorlinks,citecolor=blue,urlcolor=blue]{hyperref}

\usepackage{xr}
 \makeatletter
\newcommand*{\addFileDependency}[1]{
  \typeout{(#1)}
  \@addtofilelist{#1}
  \IfFileExists{#1}{}{\typeout{No file #1.}}
}
\makeatother

\newcommand*{\myexternaldocument}[1]{%
    \externaldocument{#1}%
    \addFileDependency{#1.tex}%
    \addFileDependency{#1.aux}%
}
\myexternaldocument{Supplementary}

\makeatletter
\newenvironment{chapquote}[2][2em]
  {\setlength{\@tempdima}{#1}%
   \def\chapquote@author{#2}%
   \parshape 1 \@tempdima \dimexpr\textwidth-2\@tempdima\relax%
   \itshape}
  {\par\normalfont\hfill \chapquote@author\hspace*{\@tempdima}\par\bigskip}
\makeatother

\renewcommand{\baselinestretch}{1.6}
\providecommand{\keywords}[1]{\textbf{\textit{Keywords: }} #1}
\newcommand{\blind}{1}

\begin{document}
	\if1\blind
	{
		\title{\bf \Large
        Tensor dynamic conditional correlation model: A new way to pursuit ``Holy Grail of investing''
		}
		\author[1]{Cheng Yu}
		\author[2]{Zhoufan Zhu}
        \author[3]{Ke Zhu}
	    \affil[1]{\it Department of Statistics and Data Science, Tsinghua University}
        \affil[2]{\it Wang Yanan Institute for Studies in Economics (WISE), Xiamen University}
        \affil[3]{\it Department of Statistics and Actuarial Science, University of Hong Kong}
		\setcounter{Maxaffil}{0}
		
		\renewcommand\Affilfont{\itshape\small}
		\date{\vspace{-5ex}}
		\maketitle
	} \fi
	\if0\blind
	{
		\bigskip
		\bigskip
		\bigskip
		\begin{center}
			{\Large\bf }
		\end{center}
		\medskip
	} \fi

\setcounter{Maxaffil}{0}
\renewcommand\Affilfont{\itshape\small}
\begin{abstract}

Style investing creates asset classes (or the so-called ``styles'') with low correlations, aligning well with the principle of ``Holy Grail of investing'' in terms of portfolio selection. The returns of styles naturally form a tensor-valued time series, which requires new tools for studying the dynamics of the conditional correlation matrix to facilitate the aforementioned principle. Towards this goal, we introduce a new tensor dynamic conditional correlation (TDCC) model, which is based on two novel treatments: trace-normalization and dimension-normalization. These two normalizations adapt to the tensor nature of the data, and they are necessary except when the tensor data reduce to vector data. Moreover, we provide an easy-to-implement estimation procedure for the TDCC model, and examine its finite sample performance by simulations. Finally, we assess the usefulness of the TDCC model in international portfolio selection across ten global markets and in large portfolio selection for 1800 stocks from the Chinese stock market.

\end{abstract}

\keywords{Holy Grail of investing; Multi-way conditional heteroskedasticity; Portfolio selection; Style investing; Tensor dynamic conditional correlation model; Tensor-valued time series}




\newpage


\section{Introduction}
\label{sec:intro}

\begin{chapquote}{\cite{dalio2017principles}}
``Having a few good uncorrelated return streams is better than having just one, and knowing how to combine return streams is even more effective than being able to choose good ones (though of course you have to do both).''
\end{chapquote}

Modern portfolio selection can be traced back to the seminal work of \cite{Markowitz1952}. For a pool of chosen assets, the related portfolio aims to diversify their unsystematic risks by optimally assigning investment weights, which depend on the conditional mean and covariance matrix of their return vector in the mean-variance framework or on the conditional covariance matrix only in the minimum-variance framework. However, the mean-variance or minimum-variance framework does not provide any clue or guidance on how to form the asset pool. With a remarkable empirical success, \cite{dalio2017principles} tackles this issue via the pursuit of his ``Holy Grail of investing'', an investing principle stating that a very effective way to make fortune is choosing an asset pool consisting of (as many as possible) assets with pairwisely uncorrelated or even negative-correlated returns. According to this investing principle, finding such a desirable asset pool becomes a primary objective before adopting the mean-variance or minimum-variance framework.

Style investing (\citealp{Barberis2003Style}) offers us a feasible way to achieve the above objective. In style investing, investors categorize assets into classes (or the so-called ``styles'') and then allocate their investment weights across these styles. For example, commonly used styles are formed in terms of the states of a single characteristic, such as ``Investment Market (or Market)'', ``Market Capitalization (or Size)'', or ``Book-to-market Ratio (or B/M)''. As demonstrated by \cite{Barberis2003Style} and other studies, many investors in financial markets pursue style investing, which increases the correlation between assets of the same style, and reduces the correlation between assets of different styles. This phenomenon gives us the financial motivation to form an asset pool consisting of styles. In other words,
we should apply the mean-variance or minimum-variance framework to assign different investment weights across styles. For the assets in each style, we do not distinguish them because of their possibly higher correlations. Therefore, in each style, we can simply invest either just one specific asset or all assets in an equally-weighted fashion.

To further make the asset pool of styles obey the principle of ``Holy Grail of investing'', we need to increase the number of styles. One natural way for this purpose is to use multiple characteristics to construct the styles, each of which is determined jointly by the states of these characteristics. Formally, we can consider $K$ different characteristics, where each $k$th characteristic has $N_k$ different states, say, state `1', state `2', $\ldots$, and state `$N_k$'. Then, the styles derived from these $K$ different characteristics naturally lead to an order-$K$ tensor time series $\mathcal{X}_t \in \mathbb{R}^{N_1 \times \cdots \times N_K}$, where the $(i_1,\ldots,i_K)$th entry of $\mathcal{X}_t$, for $1\leq i_k \leq N_k$ and $k=1,\ldots, K$, is the return of the $(i_1,\ldots,i_K)$th style, and the $(i_1,\ldots,i_K)$th style is an asset class with the $k$th characteristic having the state `$i_k$'. By construction, it is evident that the $k$th mode of $\mathcal{X}_t$ is related to the $k$th characteristic. Below, we give two illustrating examples on the construction of order-3 tensor time series $\mathcal{X}_t \in \mathbb{R}^{N_1 \times N_2 \times N_3}$:

\begin{example}\label{exa1}
(International Portfolio Selection). Constructing an international portfolio that includes assets from various markets is recognized as an effective strategy to diversify risks (\citealp{das2004systemic}; \citealp{kroencke2014international}). Inherently, ``Market'' is an obvious characteristic to construct styles. Moreover,
empirical studies in \cite{Kenneth2008Style} and \cite{Wahal2013Style} demonstrate that ``Sector'' and ``Size'' are another two important characteristics to proceed style investing in stock markets. Consequently, using the three above characteristics,
we obtain an order-3 tensor time series $\mathcal{X}_t\in \mathbb{R}^{N_1 \times N_2 \times N_3}$, where the $(i_1,i_2,i_3)$th entry of $\mathcal{X}_t$ is the return of a style that contains only one stock from market $i_1$, sector $i_2$, and with capitalization level $i_3$ at time $t$. In this case, the stocks in all the $N_1N_2N_3$ styles are drawn from $N_1$ markets (w.r.t. mode 1 of $\mathcal{X}_t$), with each market consisting of stocks from $N_2$ sectors (w.r.t. mode 2 of $\mathcal{X}_t$); moreover, the stocks in each sector are categorized into $N_3$ groups (w.r.t. mode 3 of $\mathcal{X}_t$) in terms of the market capitalization level.
\end{example}

\begin{example}\label{exa2}
(Large Portfolio Selection). As the number of assets increases rapidly in the financial market, it raises a challenging problem on how to construct large portfolios. Intuitively, style investing offers a reasonable solution for the above problem, since it is an effective way to reduce the dimension and process vast amounts of information (\citealp{mullainathan2002thinking}). Under the idea of style investing, we categorize all assets into several styles according to their state values on some characteristics. To achieve this categorization, we can utilize the knowledge of asset pricing, which leads to a great success in portfolio selection by dividing all assets into deciles according to their exposures to a specific factor. Motivated by this, the factors used for asset pricing should be good characteristics to define styles. For example, we can take three firm characteristics: ``Size'', ``Operating Profitability (OP)'', and ``Investment (Inv)'', to construct styles jointly, where these three characteristics are related to the size, profitability, and investment factors in \cite{fama2015five}. If ``Size'', ``OP'', and ``Inv'' have $N_1$, $N_2$, and $N_3$ different levels, respectively, we obtain an order-3 tensor time series  $\mathcal{X}_t\in \mathbb{R}^{N_1 \times N_2 \times N_3}$ that encompasses all those style returns, where the style return is calculated as the averaged return of all assets within each style.

Note that we do not need to consider any firm characteristic (e.g., ``B/M'') related to the value factor, since \cite{fama2015five} find that the value factor is redundant in the presence of profitability and investment factors.  Obviously, if we use some additional firm characteristics to construct styles, it would result in an even higher-order tensor time series.
\end{example}

Given an order-$K$ tensor time series $\mathcal{X}_t$, the next target is to study its conditional mean and covariance matrix for choosing the optimal investment weights across styles. Commonly, the conditional mean of $\mathcal{X}_t$ has no dynamics, as demonstrated by many financial return data. If this is not the case, we can straightforwardly explore the conditional mean of $\mathcal{X}_t$ via the tensor autoregressive (TAR) models in \cite{li2021TAR} and \cite{wang2024high}.
Therefore, due to the de-mean implementation, it suffices to assume that $\mathcal{X}_t$ has conditional mean zero with the conditional covariance matrix
\begin{equation}\label{def_ccm}
\mathbf{\Sigma}_t=E\left(\text{vec}(\mathcal{X}_t)\text{vec}(\mathcal{X}_t)'|\mathcal{F}_{t-1}\right)\in\mathbb{R}^{N\times N}
\end{equation}
for $N=\prod_{k=1}^{K} N_k$. Here,  $N$ represents the total number of styles, $\mathcal{F}_{t}=\sigma(\mathcal{X}_s; s\leq t)$ is the natural filtration containing all the available information up to time $t$, and $\text{vec}(\cdot)$ is the vectorization operator. Note that the dimension of $\mathbf{\Sigma}_t$ can be very large, even for moderate values of $N_k$. For instance, we have $N=125$ when $N_1=N_2=N_3=5$ for an order-3 $\mathcal{X}_t$.

Motivated by the seminar work of \cite{engle1982autoregressive} and \cite{bollerslev1986generalized}, one can study the dynamics of
$\mathbf{\Sigma}_t$ assuming that $\text{vec}(\mathcal{X}_t)$ follows certain multivariate generalized autoregressive conditional heteroskedasticity (GARCH) models. See the survey on GARCH-type models in \cite{bauwens2006multivariate} and \cite{francq2019garch}. However, this vectorization-based treatment is not advisable due to two drawbacks: First, it brings damage to the original tensor structure of $\mathcal{X}_t$, making it difficult to informatively
interpret the multi-way conditional heteroskedasticity with respect to $K$ different modes of $\mathcal{X}_t$; Second, it needs to estimate model parameters of order $O(N^{2})$, so the related estimation becomes computationally challenging even for moderate values of $N_k$. To circumvent these two drawbacks, \cite{yu2023matrix} propose a matrix GARCH model for the matrix-valued $\mathcal{X}_t$ (i.e., $K=2$), without implementing the vectorization operator.
This matrix GARCH model avoids an identification issue to study the two-way conditional heteroskedasticity of $\mathcal{X}_t$ via a trace-normalized vector Baba-Engle-Kraft-Kroner (BEKK) specification, which is the classical vector BEKK specification (\citealp{engle1995multivariate}) normalized by its own trace. Moreover, the matrix GARCH model
needs a key trace process to accommodate the matrix nature of $\mathcal{X}_t$, which intrinsically requires the conditional covariance matrix of each mode to have the same trace. Although this trace process captures the overall risk of all entries in $\mathcal{X}_t$, it assumes that the sum of conditional variances of all entries follows a univariate GARCH specification.
This assumption seems restrictive from a practical perspective, since it essentially requires each entry of $\mathcal{X}_t$ to have the same univariate GARCH specification for conditional variance.

\subsection{Contributions of the Current Paper}

In this paper, we find that a key identification issue exists for studying the multi-way conditional heteroskedasticity with respect to $K$ different modes of $\mathcal{X}_t$. To solve this issue, we provide a general trace-normalization solution that forms the basis of our new tensor dynamic conditional correlation (TDCC) model. The TDCC model is built on $\mathcal{X}_t$ directly rather than $\text{vec}(\mathcal{X}_t)$. It applies a similar idea as the vector DCC (VDCC) model in \cite{engle2002dynamic} to separately learn conditional variances and correlation matrices, and reduces to the VDCC model when $K=1$. Specifically, the TDCC model employs different univariate GARCH specifications for the conditional variances of the entries of $\mathcal{X}_t$, along with different VDCC-type specifications for the conditional correlation matrices of the modes of $\mathcal{X}_t$. Since the univariate GARCH specifications for conditional variances are well identified, the TDCC model automatically satisfies the intrinsic requirement that the conditional covariance matrix of each mode has the same trace. Moreover, we highlight that when $K\geq 2$, the VDCC-type specifications for the conditional correlation matrices of all modes differ from the classical VDCC specification in \cite{engle2002dynamic}, because they must account for dimension-normalization constants, generated by the devolatilized tensor of $\mathcal{X}_t$.

By dealing with conditional variances and correlation matrices separately rather than as a unified entity, the TDCC model learns $\mathbf{\Sigma}_t$ with three advantages. First, it enables the interpretation of multi-way conditional correlations with respect to $K$ different modes. Second, unlike the matrix GARCH model in \cite{yu2023matrix}, its trace process permits heterogeneity in conditional variances, since each entry of $\mathcal{X}_t$ has its own univariate GARCH specification for conditional variance. Third, it allows the use of a two-step estimation procedure to sequentially learn the conditional variances and correlation matrices, effectively reducing the computational burden of model estimation, even for the case of large $N$.
This two-step estimation procedure is similar to that in \cite{engle2002dynamic}, while enjoying an interesting difference: Owing to the tensor nature of $\mathcal{X}_t$, it has no need to estimate the intercept parameter matrices in conditional correlation matrices via the method of linear/nonlinear shrinkage (\citealp{Engle2019Large}), when dealing with most cases of large $N$.

Notably, we can also learn $\mathbf{\Sigma}_t$ by applying the VDCC model to fit $\text{vec}(\mathcal{X}_t)$. However, this treatment loses the first aforementioned advantage of the TDCC model, since it uses the vectorization operator to break the tensor structure of $\mathcal{X}_t$ for implementing a VDCC model. In addition, it only has two parameters to dynamically update the conditional correlation matrix of $\text{vec}(\mathcal{X}_t)$, whereas the TDCC model has two mode-specific parameters to update the conditional correlation matrix of each mode, adapting to the tensor nature of $\mathcal{X}_t$.

In two empirical applications, we use the TDCC model to predict $\mathbf{\Sigma}_t$ for portfolio selection. Our first application is related to Example \ref{exa1}, where the 440 styles are constructed from ten global markets, eleven sectors, with four capitalization levels. As demonstrated by the fitting results, the conditional correlations among styles are typically very small. According to the out-of-sample results, the minimum-variance portfolios selected by the TDCC-based method outperform those by other competing methods, which break the tensor structure of returns into matrix or vector counterpart during the modeling process. Hence, it indicates the importance of maintaining the tensor structure of returns when predicting $\mathbf{\Sigma}_t$.

Our second application is relevant to Example \ref{exa2}, based on 125 styles constructed from 1800 stocks according to their levels of three characteristics: ``Size'', ``OP'', and ``Inv''. Compared with the findings of the first application, the conditional correlations among these 125 styles are larger (around 0.4 on average), shedding light on the fact that styles across different markets are more likely than those from a single market to obey the principle of ``Holy Grail of investing''. In terms of out-of-sample results, our findings for the proposed mean-variance portfolios are consistent with those from the first application, further supporting the importance of the TDCC model in portfolio selection. Interestingly, by scrutinizing investment weights, we observe some clear shifts in style investing in each mode, possibly driven by some critical domestic and global shocks. In contrast, no clear style shifting appears in the first application, where the investment weights across styles in each mode are stable over time.

\subsection{Existing Works on Tensor Time Series Analysis}

Recently, there has been a growing body of statistical literature on tensor factor models;
see \cite{chen2022factor}, \cite{han2022rank}, \cite{barigozzi2023robust}, \cite{chen2024semi}, \cite{chen2024rank}, \cite{han2024tensor}, and \cite{han2021cp}, to name just a few. For $\mathcal{X}_t$ with large dimensions, the tensor factor model aims to learn its latent tensor factor (if exists) with much smaller dimensions, which is feasible for modeling and prediction based on some tensor time series models.

Till now, only few tensor time series models are available to study the conditional mean and covariance matrix $\mathbf{\Sigma}_t$ of $\mathcal{X}_t$. \cite{li2021TAR} and \cite{wang2024high} propose the TAR models to learn the conditional mean of $\mathcal{X}_t$. \cite{yu2023matrix} construct a matrix GARCH model to learn $\mathbf{\Sigma}_t$ with $K=2$. Formally, the tensor conditional mean models are the direct extensions of their vector counterparts. However, the work in \cite{yu2023matrix} shows that due to the matrix nature of data, an identification issue hinders the direct extension of vector BEKK model to their matrix counterparts. Due to the non-linearity of GARCH-type models, the extension of the VDCC model to our TDCC model is thus also not straightforward and requires noteworthy efforts to accommodate the tensor structure of $\mathcal{X}_t$, as explained above.

\subsection{Organization and Notation}

The remaining paper proceeds as follows. Section \ref{sec.model} formulates the TDCC model. Section \ref{sec: Estimation} provides a two-step estimation procedure for the TDCC model. Section \ref{sec.simulation} conducts simulation experiments. Section \ref{sec.applications} illustrates the usefulness of the TDCC model in portfolio selection. Conclusions are given in Section \ref{sec.concluding}. Appendices in the supplementary materials provide the definition of some used tensor operations, technical proofs, and some additional empirical results.


Throughout the paper, $\mathbb{R}$ is the one-dimensional Euclidean space, $\mathbf{I}_{N}$ is the $N\times N$ identity matrix, and $\mathbf{1}_N$ is the $N$-dimensional vector of ones.
For an order-$K$ tensor $\mathcal{X} \in \mathbb{R}^{N_1 \times \cdots \times N_K}$,  $\operatorname{vec}(\mathcal{X})$ and $\operatorname{mat}_{k}(\mathcal{X})$ are its vectorization and mode-k unfolding matrix, respectively; meanwhile, $\mathcal{X} \times_{k} \mathbf{A}$ is the mode-$k$ product of $\mathcal{X}$ with a matrix $\mathbf{A}\in\mathbb{R}^{M_k \times N_k}$ (see Appendix A).

\section{TDCC Model}\label{sec.model}

\subsection{General Specification and its Identification}

Consider a sequence of order-$K$ tensor-valued time series $\{\mathcal{X}_t\}_{t=1}^{T}$, where
$$\mathcal{X}_t=(x_{i_1 \cdots i_K, t})_{i_k=1,\dots, N_k} \in \mathbb{R}^{N_1 \times \cdots \times N_K}$$ and $N_k$ is a positive integer for each $k\in\{1,\dots, K\}$. Given $\mathcal{F}_{t-1}$, we capture the conditional heteroskedasticity of $\mathcal{X}_t$ via the following dynamic tensor time series model:
\begin{equation}\label{Eq2.1}
    \mathcal{X}_t = \mathcal{Z}_t \times_1 \mathbf{U}_{1t}^{1/2} \times_2 \mathbf{U}_{2t}^{1/2} \times_3 \cdots \times_K \mathbf{U}_{Kt}^{1/2}, \quad t = 1,\dots, T,
\end{equation}
where $\mathbf{U}_{kt} \in \mathcal{F}_{t-1}$ is an $N_k \times N_k$ positive definite random matrix for each $k\in\{1,\dots, K\}$,  and $\{\mathcal{Z}_t\}_{t=1}^T$ is a sequence of independent and identically distributed (i.i.d.) random tensor innovations satisfying $E\left[\text{vec}(\mathcal{Z}_t)\right] = \mathbf{0}$ and $E\left[\text{vec}(\mathcal{Z}_t)\text{vec}(\mathcal{Z}_t)' \right] = \mathbf{I}_N$.
Clearly, the general specification in \eqref{Eq2.1} is the same as that of multivariate GARCH models for vector-valued time series when $K=1$, and it allows us to study the multi-way conditional heteroskedasticity of matrix-valued or higher-order tensor-valued time series when $K\geq 2$.

For $\mathcal{X}_t$ in \eqref{Eq2.1}, we investigate its mode-$k$ conditional covariance matrix given by
\begin{equation}\label{Eq2.2}
    \mathbf{S}_{kt} = E \left[ \text{mat}_{k}(\mathcal{X}_t) \text{mat}_{k}(\mathcal{X}_t)'|\mathcal{F}_{t-1} \right]\in\mathbb{R}^{N_k\times N_k}.
\end{equation}
The study of $\mathbf{S}_{kt}$ is practically important. Regarding the order-3 tensor time series $\mathcal{X}_t$ in Example \ref{exa1},
the $(i, i)$th entry of $\mathbf{S}_{1t}$ reflects the dynamic overall risk of
 all $N_2\times N_3$ stocks from market $i$, while the $(i, j)$th entry of $\mathbf{S}_{1t}$ for $i\not=j$ reveals
 the dynamic overall conditional covariance of  all $N_2\times N_3$ stocks from market $i$ and those from market $j$.
Hence, we are able to dynamically scrutinize the market-specific risk and co-movements via $\mathbf{S}_{1t}$.
Similarly, we can dynamically explore the sector-specific (or group-specific) risk and co-movements via $\mathbf{S}_{2t}$ (or $\mathbf{S}_{3t}$) in this example. Needless to say, the above mode-specific information can largely meet the needs of investors.

Next, we give a proposition to provide the formula of $\mathbf{S}_{kt}$.

\begin{proposition}\label{pro_1}
Under \eqref{Eq2.1}, the mode-k conditional covariance matrix $\mathbf{S}_{kt}$  in \eqref{Eq2.2} satisfies:
$$\mathbf{S}_{kt}= \mathbf{U}_{kt} \prod_{\substack{l = 1 \\ l\neq k}}^{K}\operatorname{tr}(\mathbf{U}_{lt} ) \,\,\mbox{ for } k = 1, \dots, K.$$
\end{proposition}

The above result shows that we can model $K$ different matrices $\mathbf{U}_{kt}$, for
$k=1, \dots, K$, to study the dynamics of $\mathbf{S}_{kt}$.
However, since
$$ \mathbf{U}_{kt} \prod_{\substack{l = 1 \\ l\neq k}}^{K}\text{tr}(\mathbf{U}_{lt} ) = \frac{\mathbf{U}_{kt}}{a} \prod_{\substack{l = 1 \\ l\neq k}}^{K}\text{tr}(a^{\frac{1}{K-1}}\mathbf{U}_{lt})\,\,\mbox{ for any }a > 0,$$
$\mathbf{U}_{kt}$ can not be identified without any further constraints when $K\geq 2$.
Following \cite{yu2023matrix}, we tackle this identification issue by posing the following condition:
 \begin{align}\label{iden_cond}
 \text{tr}(\mathbf{U}_{kt}) = 1\,\, \mbox{ for }k = 2, \dots, K.
 \end{align}
By Proposition \ref{pro_1} and the conditions in (\ref{iden_cond}), it follows that
 \begin{align}\label{iden_cond_1}
 \mathbf{S}_{1t}=\mathbf{U}_{1t}\,\,\mbox{ and }\,\,
\mathbf{S}_{kt}=\mathbf{U}_{kt}\text{tr}(\mathbf{U}_{1t})\,\,\mbox{ for }k=2,\dots,K.
 \end{align}
Moreover, since the value of $\text{tr}(\mathbf{S}_{kt})$ is unchanged across $k$ by Proposition \ref{pro_1} and
$\text{tr}(\mathbf{S}_{1t})=\text{tr}(\mathbf{U}_{1t})$ by (\ref{iden_cond_1}), we have
$\text{tr}(\mathbf{U}_{1t})=\text{tr}(\mathbf{S}_{2t})=\cdots=\text{tr}(\mathbf{S}_{Kt})$.
Together with (\ref{iden_cond_1}), it yields
 \begin{align}\label{iden_cond_2}
 \mathbf{U}_{1t}=\mathbf{S}_{1t}\,\,\mbox{ and }\,\,
\mathbf{U}_{kt}=\frac{\mathbf{S}_{kt}}{\text{tr}(\mathbf{S}_{kt})}\,\,\mbox{ for }k=2,\dots,K.
 \end{align}
Clearly, after doing trace-normalization, the identification issue for $\mathbf{U}_{kt}$ no longer exists under the settings in (\ref{iden_cond_2}).

Notably, any modeling treatment on $\mathbf{S}_{kt}$ in (\ref{iden_cond_2}) has to satisfy the intrinsic constraints
 \begin{align}\label{general_constraints}
 \text{tr}(\mathbf{S}_{1t})=\text{tr}(\mathbf{S}_{2t})=\cdots=\text{tr}(\mathbf{S}_{Kt}) \mbox{ when } K\geq 2,
  \end{align}
due to the tensor nature of $\mathcal{X}_t$. This raises the main challenge for modeling $\mathbf{S}_{kt}$. In what follows, we tackle this challenge by decomposing $\mathbf{S}_{kt}$ into its components that pertain to the conditional variances and correlations in mode-$k$.

\subsection{Decomposition of $\mathbf{S}_{kt}$}

To model $\mathbf{S}_{kt}$ in (\ref{iden_cond_2}), we write $\mathbf{S}_{kt}=(s_{ij, kt})_{i,j=1,\dots,N_k}$ and decompose it as follows:
 \begin{align}\label{dcc_decom}
 \mathbf{S}_{kt} = \mathbf{D}_{kt} \mathbf{R}_{kt} \mathbf{D}_{kt}\,\,\mbox{ for }k=1, \dots, K,
 \end{align}
where $\mathbf{D}_{kt}=\text{diag}\{s_{11, kt}^{1/2}, \dots, s_{N_kN_k, kt}^{1/2}\}\in\mathbb{R}^{N_k\times N_k}$ is a diagonal matrix, and $\mathbf{R}_{kt}\in\mathbb{R}^{N_k\times N_k}$ is the mode-$k$ conditional correlation matrix with all entries on the main diagonal equal to one. In the sequel, we investigate $\mathbf{D}_{kt}$ and $\mathbf{R}_{kt}$ in detail.

First, in order to learn $\mathbf{D}_{kt}$, we obtain from (\ref{Eq2.2}) that
\begin{align}\label{Eq2.4}
 s_{jj,kt} &= \sum_{i_1=1}^{N_1}\cdots\sum_{i_{k-1}=1}^{N_{k-1}}\sum_{i_{k+1}=1}^{N_{k+1}}\cdots\sum_{i_K=1}^{N_K}\sigma_{i_1\cdots i_{k-1} j i_{k+1}\cdots i_K, t}^2
\end{align}
for $j=1, \dots, N_k$, where $\sigma_{i_1\cdots i_K, t}^2 = E\left[(x_{i_1\cdots i_K, t})^2|\mathcal{F}_{t-1}\right]$ is the
conditional variance of the entry $x_{i_1\cdots i_K, t}$ of $\mathcal{X}_{t}$. The result (\ref{Eq2.4}) reveals that
$\mathbf{D}_{kt}$, $k=1,\dots,K$, are only related to conditional variances. More importantly,  it follows directly from (\ref{Eq2.4}) that
\begin{align}\label{equal_trace}
\text{tr}(\mathbf{S}_{kt})=\sum_{j=1}^{N_k}s_{jj,kt}=y_t,
\end{align}
for $k=1,\dots, K$, where
\begin{align}\label{process_y_t}
y_t=\sum_{i_1=1}^{N_1}\cdots\sum_{i_K=1}^{N_K}\sigma_{i_1\cdots i_K, t}^2.
\end{align}
The result in (\ref{equal_trace}) shows that the constraints in (\ref{general_constraints}) hold if all of $\sigma_{i_1\cdots i_K, t}^2$ are well identified.

Next, in order to learn $\mathbf{R}_{kt}$, we define a devolatilized order-$K$ tensor
\begin{align}\label{def_E_tensor}
\mathcal{E}_{t}=(e_{i_1 \cdots i_K, t})_{i_k=1,\dots, N_k}\in\mathbb{R}^{N_1\times \cdots\times N_K},
\end{align}
where its $(i_1, \dots, i_K)$th entry
$e_{i_1 \cdots i_K, t}=x_{i_1\cdots i_K, t} / \sigma_{i_1\cdots i_K, t}$. Below, we offer a key result, which gives us the guide on how to model $\mathbf{R}_{kt}$.

\begin{proposition}\label{pro_2}
Under \eqref{Eq2.1}, \eqref{iden_cond_2}, and \eqref{dcc_decom}, we have
\begin{equation*}
    \frac{N_k}{N}  E \left[ \text{mat}_{k}(\mathcal{E}_t) \text{mat}_{k}(\mathcal{E}_t)'|\mathcal{F}_{t-1} \right] = \mathbf{R}_{kt} \,\,\mbox{ for } k = 1, \dots, K.
\end{equation*}
\end{proposition}

The result in Proposition \ref{pro_2} indicates that $\mathbf{R}_{kt}$, $k=1,\dots,K$, are essentially the mode-$k$ conditional correlation matrices of $\sqrt{N_k/N}\mathcal{E}_t$, where $\sqrt{N_k/N}$ is the dimension-normalization constant.
It is straightforward to see that $\sqrt{N_k/N}=1$ only when $K=1$.
In other words, we must account for this normalization effect on $\mathcal{E}_t$ when modeling $\mathbf{R}_{kt}$ for matrix-valued or higher-order tensor-valued time series with $K\geq 2$.

\subsection{Specifications of $\mathbf{D}_{kt}$ and $\mathbf{R}_{kt}$}

Based on the investigations of $\mathbf{D}_{kt}$ and $\mathbf{R}_{kt}$, we follow the idea of \cite{engle2002dynamic} to model them separately.
 Specifically, to capture the dynamics of $\mathbf{D}_{kt}$, we assume that each $\sigma_{i_1\cdots i_K, t}^2$ satisfies a scalar GARCH process (\citealp{bollerslev1986generalized}):
\begin{equation}\label{univariate_garch}
   \sigma_{i_1\cdots i_K, t}^2 = \omega_{i_1\cdots i_K} + a_{i_1\cdots i_K} x_{i_1\cdots i_K, t-1}^2 + b_{i_1\cdots i_K} \sigma_{i_1\cdots i_K, t-1}^2,
\end{equation}
where the parameters $\omega_{i_1\cdots i_K}>0$, $a_{i_1\cdots i_K}\geq 0$, and $b_{i_1\cdots i_K}\geq 0$. Consequently, by (\ref{Eq2.4}), we can summarize the specification of $\mathbf{D}_{kt}$:
\begin{align}\label{final_model_D_kt}
\begin{split}
\mathbf{D}_{kt}&=\text{diag}\big\{s_{11, kt}^{1/2}, \dots, s_{N_kN_k, kt}^{1/2}\big\},\\
 s_{jj,kt} &= \sum_{i_1=1}^{N_1}\cdots\sum_{i_{k-1}=1}^{N_{k-1}}\sum_{i_{k+1}=1}^{N_{k+1}}\cdots\sum_{i_K=1}^{N_K}\sigma_{i_1\cdots i_{k-1} j i_{k+1}\cdots i_K, t}^2,
\end{split}
\end{align}
for $j=1,\dots,N_k$, where $\sigma_{i_1\cdots i_K, t}^2$ is defined in (\ref{univariate_garch}). Since all $\sigma_{i_1\cdots i_K, t}^2$ are well identified under (\ref{univariate_garch}), we know that the constraints in (\ref{general_constraints}) hold with $\text{tr}(\mathbf{S}_{kt})=y_t$ for $k=1,\dots, K$, where $y_t$ in (\ref{process_y_t}) is essentially a trace process that captures the overall risk of all entries in $\mathcal{X}_t$.

Moreover, to capture the dynamics of $\mathbf{R}_{kt}$, we consider the following VDCC-type specification for mode-specific conditional correlation matrix:
\begin{equation}\label{Eq2.10}
    \mathbf{R}_{kt} = \mathbf{Q}_{kt,*}^{-1} \mathbf{Q}_{kt}  \mathbf{Q}_{kt,*}^{-1},
\end{equation}
where $\mathbf{Q}_{kt}=(q_{ij, kt})_{i,j=1,\dots,N_k}$ satisfying
\begin{equation}\label{Eq2.8}
    \mathbf{Q}_{kt} = (1 - \alpha_k - \beta_k) \mathbf{C}_k + \alpha_k \Big[\frac{N_k}{N} \text{mat}_{k}(\mathcal{E}_{t-1}) \text{mat}_{k}(\mathcal{E}_{t-1})'\Big]  + \beta_k \mathbf{Q}_{k,t-1},
\end{equation}
with the parameters $\alpha_k\geq 0$, $\beta_k\geq 0$, $\alpha_k+\beta_k<1$, and the intercept parameter matrix 
\begin{equation}\label{Eq2.12}
\mathbf{C}_k = E(\mathbf{R}_{kt})=\frac{N_k}{N}  E \left[ \text{mat}_{k}(\mathcal{E}_t) \text{mat}_{k}(\mathcal{E}_t)'\right],
\end{equation}
and $\mathbf{Q}_{kt,*}=\text{diag}\{q_{11, kt}^{1/2}, \dots, q_{N_kN_k, kt}^{1/2}\}$ is an $N_k\times N_k$ diagonal matrix.
Clearly, the process $\mathbf{Q}_{kt,*}$ in (\ref{Eq2.10}) is to ensure that
all entries on the main diagonal line of $\mathbf{R}_{kt}$ are equal to one, and
the process $\mathbf{Q}_{kt}$ in (\ref{Eq2.8}) is motivated by Proposition \ref{pro_2} and the variance-targeting method (\citealp{Engle2019Large}) to investigate the dynamics of $\mathbf{R}_{kt}$.

In (\ref{Eq2.10})--(\ref{Eq2.12}), the condition of $E(\mathbf{R}_{kt}) = E(\mathbf{Q}_{kt})$ is implicitly assumed for the setting of $\mathbf{C}_k = E(\mathbf{R}_{kt})$. From a theoretical viewpoint, the failure of this condition tends to cause an inconsistent estimation issue of $\mathbf{C}_k$, which can be solved using the corrected version of $\mathbf{R}_{kt}$ in \cite{aielli2013dynamic}. However, from a practical viewpoint, the failure of this condition is harmless. Indeed, as argued in \cite{Engle2019Large}, we commonly have $\mathbf{Q}_{kt,*}\approx \mathbf{I}_{N_k}$ in applications, so we follow the convention in the literature to model $\mathbf{R}_{kt}$ without doing correction.

\subsection{Specification of TDCC Model}

Under the general specification of $\mathcal{X}_t$ in (\ref{Eq2.1}), the identification requirements of $\mathbf{U}_{kt}$ in (\ref{iden_cond_2}), and the decomposition of $\mathbf{S}_{kt}$ in (\ref{dcc_decom}), we give the specification of our TDCC model as follows:
\begin{align}\label{tdcc_model}
\begin{split}
    \mathcal{X}_t & = \mathcal{Z}_t \times_1 \mathbf{U}_{1t}^{1/2} \times_2 \mathbf{U}_{2t}^{1/2} \times_3 \cdots \times_K \mathbf{U}_{Kt}^{1/2},\\
    \mathbf{U}_{1t} &= \mathbf{D}_{1t} \mathbf{R}_{1t} \mathbf{D}_{1t},\\
    \mathbf{U}_{kt} &= \mathbf{D}_{kt} \mathbf{R}_{kt} \mathbf{D}_{kt} / y_t,  \text{ for } k = 2, \dots, K,
\end{split}
\end{align}
where $y_t$ is defined in (\ref{process_y_t}), and $\mathbf{D}_{kt}$ and $\mathbf{R}_{kt}$, $k=1,\dots, N_k$, are defined in (\ref{final_model_D_kt}) and (\ref{Eq2.10}), respectively.
Under the TDCC model in (\ref{tdcc_model}), the mode-$k$ conditional correlation matrix of
$\mathcal{X}_t$ is $\mathbf{R}_{kt}$ and mode-$k$ conditional covariance matrix of $\mathcal{X}_t$ is
$$\mathbf{S}_{kt}=\left\{
\begin{array}{ll}
\mathbf{U}_{1t}, & \mbox{ when } k=1; \\
y_t\mathbf{U}_{kt}, & \mbox{ when }k=2,\dots, K.
\end{array}\right.$$

When $K=1$, the TDCC model has a linkage to the existing scalar or vector GARCH-type models. Specifically, when $K=1$ and $N_1 > 1$, we have $\mathbf{U}_{1t} = \mathbf{D}_{1t}\mathbf{R}_{1t}\mathbf{D}_{1t}$, so the TDCC model reduces to the VDCC model in \cite{engle2002dynamic}. When $K = 1$ and $N_1 = 1$, the TDCC model further reduces to the scalar GARCH model in \cite{bollerslev1986generalized}.

When $K=2$, the TDCC model offers a new way to study the two-way conditional heteroskedasticity of matrix-valued time series. In the literature, \cite{yu2023matrix} make the only attempt towards this goal by proposing a matrix GARCH model. For a matrix-valued time series $\mathcal{X}_t\in \mathbb{R}^{N_1 \times N_2}$, the matrix GARCH model handles the constraint in (\ref{general_constraints}) by assuming
\begin{align}\label{matrix_garch_model}
\mathbf{S}_{1t}&=\frac{\mathbf{S}_{1t}^{*}}{\text{tr}(\mathbf{S}_{1t}^{*})}y_t\,\,\, \mbox{ and }\,\,\,\mathbf{S}_{2t}=\frac{\mathbf{S}_{2t}^{*}}{\text{tr}(\mathbf{S}_{2t}^{*})}y_t,
\end{align}
where $y_t$ is the trace process defined in (\ref{process_y_t}). Clearly, under (\ref{matrix_garch_model}) we have $\text{tr}(\mathbf{S}_{1t})=\text{tr}(\mathbf{S}_{2t})=y_t$, so the constraint in (\ref{general_constraints}) holds automatically, regardless of the specifications of $\mathbf{S}_{1t}^{*}$, $\mathbf{S}_{2t}^{*}$, and $y_t$.
For the matrix GARCH model, it further assumes
\begin{align}
\mathbf{S}_{kt}^{*} &= \mathbf{A}_{k0} \mathbf{A}_{k0}' + \mathbf{A}_{k1} [\text{mat}_{k}(\mathcal{X}_{t-1}) \text{mat}_{k}(\mathcal{X}_{t-1})'] \mathbf{A}_{k1}' + \mathbf{A}_{k2} \mathbf{S}_{k,t-1}^{*} \mathbf{A}_{k2}', \label{bekk_model}\\
y_t &= \omega + \alpha\, \text{tr}\big(\text{vec}(\mathcal{X}_{t-1}) \text{vec}(\mathcal{X}_{t-1})'\big) + \beta y_{t-1}, \label{garch_model}
\end{align}
where $\mathbf{A}_{k0}$, $\mathbf{A}_{k1}$, and $\mathbf{A}_{k2}$ are $N_k\times N_k$ parameter matrices,  $\mathbf{A}_{k0}$ are lower triangular matrices with the $(1,1)$th entry being one and the other diagonal entries being non-negative, and the scalar parameters $\omega > 0$, $\alpha \ge 0$, and $\beta \ge 0$ ensure the positivity of $y_t$. In view of (\ref{Eq2.2}), the BEKK specification (\citealp{engle1995multivariate}) of $\mathbf{S}_{kt}^{*}$ in (\ref{bekk_model}) is motivated to capture the mode-$k$ conditional heteroskedasticity.
Meanwhile, since $y_t=\text{tr}\big[E(\text{vec}(\mathcal{X}_{t}) \text{vec}(\mathcal{X}_{t})'|\mathcal{F}_{t-1})\big]$, the scalar GARCH specification (\citealp{bollerslev1986generalized}) in (\ref{garch_model}) is given for $y_t$. Although both TDCC (with $K=2$) and matrix GARCH models make use of $y_t$, they capture the dynamics of $y_t$ in two different ways. The TDCC model allows each $\sigma_{i_1i_{2},t}^2$ (within $y_t$) to have its own GARCH specification, whereas the matrix GARCH model essentially assumes all of $\sigma_{i_1i_{2},t}^2$ to have the same scalar GARCH specification: $\sigma_{i_1i_{2},t}^2=\omega+\alpha x_{i_1i_2,t-1}^2+\beta \sigma_{i_1i_{2},t-1}^2$.
From an empirical viewpoint, the style returns most likely do not share the same specification of their conditional variances, so the way to capture $y_t$ in the TDCC model seems more appropriate. Notably, this advantage of the TDCC model remains when the matrix GARCH model is extended to its tensor counterpart with $K\geq 3$.

\subsection{Exploration of $\mathbf{\Sigma}_t$ via TDCC Model}

For $\mathbf{\Sigma}_t=(\sigma^2_{i_1 \cdots i_K, t})_{i_k=1,\dots, N_k}$ in (\ref{def_ccm}), we can directly write it as
\begin{align}\label{sigma_tdcc}
    \mathbf{\Sigma}_t = \mathbf{D}_t \mathbf{R}_t \mathbf{D_t},
\end{align}
where $\mathbf{D}_t = \text{diag}\{\sigma_{1\cdots 1,t}, \dots, \sigma_{N_1\cdots N_K, t}\}\in\mathbb{R}^{N\times N}$ is a diagonal matrix, and
$\mathbf{R}_t\in\mathbb{R}^{N\times N}$ is the conditional correlation matrix of $\text{vec}(\mathcal{X}_t)$ with all entries on the main diagonal equal to one. Indeed, using the fact that
\begin{align}\label{e_t_x_t}
\operatorname{vec}(\mathcal{E}_t) = \mathbf{D}_t^{-1} \operatorname{vec}(\mathcal{X}_t)
\end{align}
with $\mathcal{E}_t$ defined in (\ref{def_E_tensor}), we can easily show
\begin{align}\label{explain_R_t}
\begin{split}
E\big[\operatorname{vec}(\mathcal{E}_t)\operatorname{vec}(\mathcal{E}_t)'|\mathcal{F}_{t-1}\big]
& =\mathbf{D}_t^{-1} E\big[\operatorname{vec}(\mathcal{X}_t)\operatorname{vec}(\mathcal{X}_t)'|\mathcal{F}_{t-1}\big] \mathbf{D}_t^{-1} = \mathbf{D}_t^{-1} \mathbf{\Sigma}_t \mathbf{D}_t^{-1}  = \mathbf{R}_t,
\end{split}
\end{align}
where the last equality holds by (\ref{sigma_tdcc}). Because $\mathcal{E}_t$ is the devolatilized counterpart of $\mathcal{X}_t$, result (\ref{explain_R_t}) matches our expectation.

Under (\ref{Eq2.1}), we have
\begin{align}\label{Eq2.13}
\begin{split}
    \text{vec}(\mathcal{X}_t) & = \Big(\bigotimes_{k=K}^{1}\mathbf{U}_{kt}\Big)^{1/2} \text{vec}(\mathcal{Z}_t).
\end{split}
\end{align}
Since $E\left[\text{vec}(\mathcal{Z}_t)\right] = \mathbf{0}$ and $E\left[\text{vec}(\mathcal{Z}_t)\text{vec}(\mathcal{Z}_t)' \right] = \mathbf{I}_N$, by \eqref{iden_cond_2}, \eqref{dcc_decom}, and (\ref{Eq2.13}), it follows that
\begin{align}\label{sigma_exp}
\begin{split}
\mathbf{\Sigma}_t  = \bigotimes_{k=K}^{1}\mathbf{U}_{kt}
= \Big( y_t ^{-\frac{K-1}{2}}\bigotimes_{k=K}^{1}\mathbf{D}_{kt} \Big) \Big( \bigotimes_{k=K}^{1}\mathbf{R}_{kt}  \Big) \Big( y_t ^{-\frac{K-1}{2}}\bigotimes_{k=K}^{1}\mathbf{D}_{kt} \Big).
\end{split}
\end{align}
Clearly, $y_t ^{-(K-1)/2}\bigotimes_{k=K}^{1}\mathbf{D}_{kt}\in\mathbb{R}^{N\times N}$ is a diagonal matrix with all positive entries on the main diagonal, and $\bigotimes_{k=K}^{1}\mathbf{R}_{kt}\in\mathbb{R}^{N\times N}$ is a matrix with all entries on the main diagonal equal to one. Hence, by (\ref{sigma_tdcc}), we know that
under (\ref{Eq2.1}), \eqref{iden_cond_2}, and \eqref{dcc_decom},
\begin{align}\label{D_and_R}
\mathbf{D}_t  = y_t ^{-\frac{K-1}{2}}\bigotimes_{k=K}^{1}\mathbf{D}_{kt}\quad \mbox{ and }\quad
\mathbf{R}_t = \bigotimes_{k=K}^{1}\mathbf{R}_{kt},
\end{align}
leading to the learning of $\mathbf{\Sigma}_t$ in (\ref{sigma_tdcc}) by the TDCC model.

Besides the use of TDCC model for $\mathcal{X}_t$, one can also study $\mathbf{\Sigma}_t$ in (\ref{def_ccm}) via a VDCC model for $\text{vec}(\mathcal{X}_t)$. Formally, the VDCC model is just a special TDCC model with $K=1$, and it assumes in (\ref{sigma_tdcc}) that $\sigma^2_{i_1 \cdots i_K, t}$ in $\mathbf{D}_t$ has a GARCH specification as in (\ref{univariate_garch}), and
\begin{align}\label{vector_dcc_model}
\mathbf{R}_t =   [\mathbf{Q}_{t,*}^{\dag}]^{-1} \mathbf{Q}_{t}^{\dag}  [\mathbf{Q}_{t,*}^{\dag}]^{-1},
\end{align}
where $\mathbf{Q}_{t}^{\dag}=(q_{ij, t}^{\dag})_{i,j=1,\dots,N}$ satisfying
\begin{align}\label{vdcc_spe}
\mathbf{Q}_t^{\dag} = (1 - \alpha^{\dag} - \beta^{\dag})\mathbf{C}^{\dag} + \alpha^{\dag} \text{vec}(\mathcal{E}_{t-1}) \text{vec}(\mathcal{E}_{t-1})' + \beta^{\dag} \mathbf{Q}_{t-1}^{\dag},
\end{align}
with the parameters $\alpha^{\dag}\geq 0$, $\beta^{\dag}\geq 0$, $\alpha^{\dag}+\beta^{\dag}<1$, and the intercept parameter matrix $\mathbf{C}^{\dag} = E\big[ \text{vec}(\mathcal{E}_t) \text{vec}(\mathcal{E}_t)' \big]$, and $\mathbf{Q}_{t,*}^{\dag} =\text{diag}\big\{q_{11, t}^{\dag 1/2}, \dots, q_{NN, t}^{\dag 1/2}\big\}$ is an $N\times N$ diagonal matrix.

Clearly, the TDCC and VDCC models have the same way of specifying $\mathbf{D}_t$. Thus, both of them are more appropriate than the matrix GARCH model (or its tensor counterpart) in terms of the investigation of conditional variances. Moreover, we notice from (\ref{D_and_R}) and (\ref{vector_dcc_model}) that the TDCC model captures $\mathbf{R}_t$ differently from the VDCC model to gain two advantages. First, the TDCC model maintains the original tensor structure of $\mathcal{X}_t$ without using the vectorization operator, so it allows for the interpretation of mode-specific conditional correlations via $\mathbf{R}_{kt}$. Second, the TDCC model has the mode-specific parameters $\alpha_k$ and $\beta_k$, offering a remarkable flexibility to fit tensor-valued time series. In contrast, the VDCC model has only two parameters, $\alpha^{\dag}$ and $\beta^{\dag}$, to update the dynamics of $\mathbf{R}_t$ for learning $\mathbf{\Sigma}_t$. 

\section{Estimation of TDCC Model}\label{sec: Estimation}

This section provides a two-step estimation procedure for the TDCC model in (\ref{tdcc_model}). Let $\theta = (\eta', \phi')'$ denote the parameters to estimate, where $\eta$ denotes the parameters of all  $\mathbf{D}_k$ in (\ref{final_model_D_kt}), and $\phi$ denotes the additional parameters of all $\mathbf{R}_k$ in (\ref{Eq2.10}). Assume that $\mathcal{Z}_t$ in (\ref{tdcc_model}) follows a standard tensor normal distribution $\text{TN}_{N_1, N_2, \dots, N_K}(\mathbf{0}, \mathbf{I}_{N_1}, \dots, \mathbf{I}_{N_K})$ in $\mathbb{R}^{N_1 \times \cdots \times N_K}$. Then, by (\ref{tdcc_model})  we have
$\mathcal{X}_t | \mathcal{F}_{t-1} \sim \text{TN}_{N_1, N_2, \dots, N_K}\left(\mathbf{0}, \mathbf{U}_{1t}, \mathbf{U}_{2t}, \dots, \mathbf{U}_{Kt}\right),$ which entails
\begin{equation}\label{dist_vec_x}
    \text{vec}(\mathcal{X}_t) | \mathcal{F}_{t-1} \sim \text{N}_{N}\big(\mathbf{0}, \mathbf{\Sigma}_t\big).
\end{equation}
Here, $\mathbf{\Sigma}_t=\bigotimes_{k=K}^{1}\mathbf{U}_{kt}=\mathbf{D}_t\mathbf{R}_t\mathbf{D}_t$ by (\ref{sigma_exp})--(\ref{D_and_R}), and $\text{N}_{k}(\mathbf{0}, \mathbf{\Omega})$ stands for the  multivariate normal distribution in $\mathbb{R}^{k}$ with mean vector $\mathbf{0}$ and variance-covariance matrix $\mathbf{\Omega}$.

By (\ref{dist_vec_x}), it is straightforward to see that the log-likelihood function (ignoring constants) of $\{\text{vec}(\mathcal{X}_1),\dots, \text{vec}(\mathcal{X}_T)\}$ from the TDCC model is
\begin{align}\label{likelihood_func}
\begin{split}
    L_T(\theta) =&  -\frac{1}{2T}\sum_{t=1}^T \left\{ \log|\mathbf{\Sigma}_t| + \text{vec}(\mathcal{X}_t)'\mathbf{\Sigma}_t^{-1}\text{vec}(\mathcal{X}_t)  \right\}   \\
    =& -\frac{1}{2T}\sum_{t=1}^T \left\{\log|\mathbf{D}_t|^{2} + \text{vec}(\mathcal{X}_t)' \mathbf{D}_t^{-2} \text{vec}(\mathcal{X}_t)\right\}  \\
    &-\frac{1}{2T}\sum_{t=1}^T \left\{ \log| \mathbf{R}_t | + \text{vec}(\mathcal{E}_t)' \mathbf{R}_t^{-1}  \text{vec}(\mathcal{E}_t) - \text{vec}(\mathcal{E}_t)' \text{vec}(\mathcal{E}_t) \right\} \\
    =&: L_v(\eta) + L_c(\eta, \phi),
\end{split}
\end{align}
where the second equality holds by (\ref{e_t_x_t}), $L_v(\eta)$ is the volatility term with
\begin{align} \label{EqLv}
\begin{split}
    L_v(\eta) & = -\frac{1}{2T}\sum_{t=1}^T \left\{\log|\mathbf{D}_t|^2 + \text{vec}(\mathcal{X}_t)' \mathbf{D}_t^{-2} \text{vec}(\mathcal{X}_t)\right\}\\
    & = \sum_{i_1=1}^{N_1}\cdots \sum_{i_K=1}^{N_K} \Big\{-\frac{1}{2T}\sum_{t=1}^{T}\Big( \log(\sigma_{i_1\cdots i_K,t}^2 ) + \frac{ (\mathcal{X}_{t})_{i_1\cdots i_K}^2}{\sigma_{i_1\cdots i_K,t}^2}\Big)\Big\},
\end{split}
\end{align}
and  $L_c(\eta, \phi)$ is the correlation term with
\begin{align}  \label{EqLc}
\begin{split}
     L_c(\eta, \phi) &= -\frac{1}{2T}\sum_{t=1}^T \left\{ \log| \mathbf{R}_t | + \text{vec}(\mathcal{E}_t)' \mathbf{R}_t^{-1}  \text{vec}(\mathcal{E}_t) - \text{vec}(\mathcal{E}_t)' \text{vec}(\mathcal{E}_t) \right\} \\
     &= -\frac{1}{2T}\sum_{t=1}^T \Big\{ \sum_{k=1}^{K}\frac{N}{N_k}\log| \mathbf{R}_{k,t} | + \text{vec}(\mathcal{E}_t)' \Big(\bigotimes_{k=K}^{1}\mathbf{R}_{k,t}^{-1}\Big)  \text{vec}(\mathcal{E}_t) - \text{vec}(\mathcal{E}_t)' \text{vec}(\mathcal{E}_t) \Big\}.
\end{split}
\end{align}
Note that $\mathbf{D}_t$ and $\mathcal{E}_t$ depend on $\eta$, and $\mathbf{R}_{k,t}$ depends on $\eta$ and $\phi$. For notational convenience, we have suppressed these arguments for ease of presentation.

As in \cite{engle2002dynamic}, we next give a two-step estimation procedure to estimate $\theta$:

\textit{Step 1:} Compute $\widehat{\eta} = \argmax_{\eta} L_v(\eta)$;

\textit{Step 2:} Compute $\widehat{\phi} = \argmax_{\phi} L_{c}(\widehat{\eta}, \phi)$.

\noindent Combining $\widehat{\eta}$ and $\widehat{\phi}$, we obtain our estimator $\widehat{\theta}=(\widehat{\eta}', \widehat{\phi}')'$ for $\theta$. It is evident that we estimate
$\eta$ and $\phi$ step-wisely rather than jointly. The reason is that a joint optimization of $L_T(\theta)$ in (\ref{likelihood_func}) is computationally challenging due to the high dimension of $\theta$. By estimating $\theta$ through two small tasks in Steps 1 and 2, we can largely relieve the computational burden.

Specifically, since $L_{v}(\eta)$ in (\ref{EqLv}) is essentially the sum of $N$ different individual GARCH likelihoods, we optimize each individual GARCH likelihood separately to obtain the estimators of $\omega_{i_1\cdots i_K}$, $a_{i_1\cdots i_K}$, and $b_{i_1\cdots i_K}$ in (\ref{univariate_garch}). Then, we combine all the estimators of $\omega_{i_1\cdots i_K}$, $a_{i_1\cdots i_K}$, and $b_{i_1\cdots i_K}$ from these GARCH likelihoods to form $\widehat{\eta}$. Next,
we write $\phi=(\phi_1',\phi_2')'$, where $\phi_1$ contains all the parameters consisting of $\mathbf{C}_k$ in (\ref{Eq2.12}) for $k=1,\dots, K$, and $\phi_1$ includes all the parameters $\alpha_k$ and $\beta_k$ in (\ref{Eq2.8}) for $k=1,\dots, K$. Based on $\widehat{\eta}$, we calculate the estimated devolatilized return series $\widehat{\mathcal{E}}_t$. Using $\widehat{\mathcal{E}}_t$, we estimate $\mathbf{C}_k$ by its sample version
\begin{align}\label{new_est_C}
\widehat{\mathbf{C}}_k=\frac{1}{T}\sum_{t=1}^{T}\Big[\frac{N_k}{N} \text{mat}_{k}(\widehat{\mathcal{E}}_t)\text{mat}_k(\widehat{\mathcal{E}}_t)'\Big],
\end{align}
leading to the estimator $\widehat{\phi}_1$ for $\phi_1$. Moreover, we calculate $\widehat{\phi}_2$, the estimator of $\phi_2$, by optimizing $L_{c}(\widehat{\eta}, \widehat{\phi}_1, \phi_2)$, which is defined in the same way as $L_c(\eta, \phi)$ in (\ref{EqLc}) with $\mathcal{E}_t$ and $\mathbf{C}_k$ replaced by $\widehat{\mathcal{E}}_t$ and $\widehat{\mathbf{C}}_k$, respectively. The optimization of $L_{c}(\widehat{\eta}, \widehat{\phi}_1, \phi_2)$ is computationally easy, since $\phi_2$ only has $2K$ different parameters. Finally, we stack $\widehat{\phi}_1$ and $\widehat{\phi}_2$ together to get the estimator $\widehat{\phi}=(\widehat{\phi}_1',\widehat{\phi}_2')'$.

Notably, we learn $\mathbf{C}_k$ using the sample covariance estimator $\widehat{\mathbf{C}}_k$ in (\ref{new_est_C}), rather than the linear shrinkage estimator \citep{ledoit2004well} and the non-linear shrinkage estimator \citep{ledoit2012nonlinear}. This is different from the treatment for the VDCC model in \cite{Engle2019Large}. In the VDCC model, the sample covariance estimator of $\mathbf{C}^{\dag}$ in (\ref{vdcc_spe}) is
$$\widehat{\mathbf{C}}^{\dag} = \frac{1}{T}\sum_{t=1}^{T}\operatorname{vec}(\widehat{\mathcal{E}}_t) \operatorname{vec}(\widehat{\mathcal{E}}_t)',$$
where $\widehat{\mathcal{E}}_t$ is an estimator of $\mathcal{E}_t$, and $\operatorname{vec}(\widehat{\mathcal{E}}_t)\in\mathbb{R}^{N}$. Even though $\widehat{\mathcal{E}}_t$ is consistent to $\mathcal{E}_t$, it is well known that $\widehat{\mathbf{C}}^{\dag}$ is a poor estimator for $\mathbf{C}^{\dag}$, when $N/T \rightarrow c \in (0, \infty)$; see, for example, \cite{lam2020high}. To overcome this inconsistency issue in the case of large $N$, the linear/nonlinear shrinkage estimators are thus needed for learning $\mathbf{C}^{\dag}$ (\citealp{Engle2019Large}). Interestingly, under the TDCC model, we do not encounter this issue in most cases. To see it, we re-write
\begin{align}\label{explain_C}
    \widehat{\mathbf{C}}_k= \frac{1}{TN_{-k}} \sum_{t=1}^{T}\sum_{i=1}^{N_{-k}} \operatorname{mat}_{k}(\widehat{\mathcal{E}}_t)_{[i]} \operatorname{mat}_{k}(\widehat{\mathcal{E}}_t)_{[i]}',
\end{align}
where $N_{-k}=N/N_{k}$ and $\operatorname{mat}_{k}(\widehat{\mathcal{E}}_t)_{[i]} \in \mathbb{R}^{N_k}$ is the $i$th column of $\operatorname{mat}_{k}(\widehat{\mathcal{E}}_t)$ for $i = 1, \dots, N_{-k}$. Formula (\ref{explain_C}) clearly indicates that ``effective'' sample size to propose $\widehat{\mathbf{C}}_k$ is $T N_{-k}$ instead of $T$. Hence,
$\widehat{\mathbf{C}}_k$ is a good estimator of $\mathbf{C}_k$, when $\widehat{\mathcal{E}}_t$ is close to $\mathcal{E}_t$ and
\begin{align}\label{condition_C_k}
\frac{N_{k}}{TN_{-k}}\to 0,
\end{align}
where $N_{k}$ is the dimension of $\operatorname{mat}_{k}(\widehat{\mathcal{E}}_t)_{[i]}$. For most of tensor time series, condition (\ref{condition_C_k}) is mild, making it unnecessary to use the linear/nonlinear shrinkage estimators for $\mathbf{C}_k$ (see the related numerical evidence in Section \ref{sec.simulation}).

\section{Simulations}\label{sec.simulation}

\subsection{Simulation Setting}
We generate 1000 replications of the tensor-valued time series $\{\mathcal{X}_t\}_{t=1}^{T}$ from the TDCC model in (\ref{tdcc_model}) with $K=3$, $N_1=10$, $N_2=11$, and $N_3=4$, where $\{\mathcal{Z}_t\}_{t=1}^{T}$ are i.i.d. random tensor errors from the standard tensor normal distribution $\text{TN}_{N_1, N_2, N_3}(\mathbf{0}, \mathbf{I}_{N_1}, \mathbf{I}_{N_2}, \mathbf{I}_{N_3})$, and the sample size $T = 500$, $750$, and $1000$ that correspond to approximately $2$-year, $3$-year, and $4$-year daily data. For the true parameters of the scalar volatility process in (\ref{univariate_garch}), we take $\omega_{i_1\cdots i_K} = 0.4$, $a_{i_1\cdots i_K} = 0.05$, and $b_{i_1\cdots i_K} = 0.9$. For the true parameters of the conditional correlation matrix process in (\ref{Eq2.10}), we set $\alpha_k  = 0.05$, $\beta_k = 0.93$, and $\mathbf{C}_{k}=\mathbf{C}_{k}^{est}$ for $k=1,2,3$. Here, our way of selecting true parameters is similar to \cite{Engle2019Large}, and $\mathbf{C}_{k}^{est}$ is the estimated value of $\mathbf{C}_k$, based on the global daily stock indexes data from 2020/05/03 to 2022/12/30 in Section \ref{sec_5_1}.

For each replication, we fit $\{\mathcal{X}_t\}_{t=1}^{T}$ via different candidate methods, each of which is used to calculate $\widehat{\mathbf{\Sigma}}_t$ (i.e., the estimator of $\mathbf{\Sigma}_t$). To quantify the learning performance of $\mathbf{\Sigma}_t$, we follow \cite{Engle2019Large} to consider the loss function
\begin{align}\label{Loss}
\begin{split}
    L &= \frac{1}{T}\sum_{t=1}^{T}\mathcal{L}(\widehat{\mathbf{\Sigma}}_t, \mathbf{\Sigma}_t),
\end{split}
\end{align}
where
\begin{align*}
   \mathcal{L}(\widehat{\mathbf{\Sigma}}, \mathbf{\Sigma}) = \frac{\text{tr}(\widehat{\mathbf{\Sigma}}^{-1}\mathbf{\Sigma}\widehat{\mathbf{\Sigma}}^{-1}) / N }{[ \text{tr}(\widehat{\mathbf{\Sigma}}^{-1})/ N ]^2} - \frac{1}{\text{tr}(\mathbf{\Sigma}^{-1}) / N }
\end{align*}
for $N = N_1 \times N_2 \times N_3 = 440$.

\subsection{List of Candidate Methods}\label{sec_4_2_candidate}

We consider nine candidate methods for the purpose of comparison. Methods 1--3 are based on the TDCC model, but with different ways to estimate $\mathbf{C}_k$. Methods 4--5  are derived from the matrix DCC (MDCC) model (i.e., the TDCC model with $K=2$), which is applied to $\text{mat}_k(\mathcal{X}_t)$. Methods 6--9  pertain to the VDCC model that is implemented on $\text{vec}(\mathcal{X}_t)$. See their details below:

\begin{itemize}
\item[1.] TDCC-S: TDCC model for $\mathcal{X}_t$ with $\mathbf{C}_1$, $\mathbf{C}_2$, and $\mathbf{C}_3$ estimated by the sample covariance matrix method as in (\ref{new_est_C}).

\item[2.] TDCC-LS: TDCC model for $\mathcal{X}_t$ with $\mathbf{C}_1$, $\mathbf{C}_2$, and $\mathbf{C}_3$ estimated by the linear shrinkage method in \cite{ledoit2004well}.

\item[3.] TDCC-NLS: TDCC model for $\mathcal{X}_t$ with $\mathbf{C}_1$, $\mathbf{C}_2$, and $\mathbf{C}_3$ estimated by the nonlinear shrinkage method in \cite{ledoit2012nonlinear}.

\item[4.] MDCC1-NLS: MDCC model for $\text{mat}_1(\mathcal{X}_t)$ with the corresponding intercept parameter matrices $\mathbf{C}_1^{\ddag}$ and $\mathbf{C}_2^{\ddag}$ estimated by the nonlinear shrinkage method.

\item[5.] MDCC2-NLS: MDCC model for $\text{mat}_2(\mathcal{X}_t)$ with the corresponding intercept parameter matrices $\mathbf{C}_1^{\ddag}$ and $\mathbf{C}_2^{\ddag}$ estimated by the nonlinear shrinkage method.

\item[6.] MDCC3-NLS: MDCC model for $\text{mat}_3(\mathcal{X}_t)$ with the corresponding intercept parameter matrices $\mathbf{C}_1^{\ddag}$ and $\mathbf{C}_2^{\ddag}$ estimated by the nonlinear shrinkage method.

\item[7.] VDCC-S: VDCC model for $\text{vec}(\mathcal{X}_t)$ with the corresponding intercept parameter matrix  $\mathbf{C}^{\dag}$ estimated by the sample covariance matrix method.

\item[8.] VDCC-LS: DCC model for $\text{vec}(\mathcal{X}_t)$  with the corresponding intercept parameter matrix  $\mathbf{C}^{\dag}$ estimated by the linear shrinkage method.

\item[9.] VDCC-NLS: DCC model for $\text{vec}(\mathcal{X}_t)$  with the corresponding intercept parameter matrix  $\mathbf{C}^{\dag}$ estimated by the nonlinear shrinkage method.
\end{itemize}

\subsection{Results}

Based on the results of 1000 replications, Table \ref{Table_TN} reports the average loss in (\ref{Loss}) for nine different candidate methods. From this table, we find that Methods 1--3 based on the TDCC model perform much better than other methods. This aligns with our expectation, indicating the importance of retaining the tensor data structure for learning $\mathbf{\Sigma}_t$. Moreover, we observe from Methods 1--3 that the estimation of $\mathbf{C}_k$ using each method leads to almost the same loss. This observation validates our statements at the end of Section \ref{sec: Estimation}. As demonstrated by the results of Methods 7--9, the use of linear shrinkage and nonlinear shrinkage methods is necessary for the high-dimensional VDCC model. This is consistent to the finding in \cite{Engle2019Large}.

\begin{table}[htbp]
  \centering
  \small
  \caption{Average loss for nine different methods}
    \begin{tabular}{ccccccc}
    \toprule
    $T$     &       & TDCC-S & TDCC-LS & TDCC-NLS & MDCC1-NLS & MDCC2-NLS \\
    \midrule
    500   &       & 0.150  & 0.149  & 0.148  & 0.658  & 0.651  \\
    750   &       & 0.096  & 0.096  & 0.095  & 0.578  & 0.572  \\
    1000  &       & 0.079  & 0.079  & 0.078  & 0.554  & 0.546  \\
    \midrule
    $T$     &       & MDCC3-NLS & DCC-S & DCC-LS & DCC-NLS &  \\
    \cmidrule{1-6}
    500   &       & 2.099  & 17.658 & 2.645 & 2.503 &  \\
    750   &       & 1.950  & 10.013 & 2.424 & 2.265 &  \\
    1000  &       & 1.763  & 6.129 & 2.281 & 2.133 &  \\
    \bottomrule
    \end{tabular}%
   \label{Table_TN}%
\end{table}%

\section{Applications}\label{sec.applications}
This section gives two applications to show the usefulness of our TDCC model in portfolio selection. The data for the first and second applications are collected fromYahoo Finance (\url{www.finance.yahoo.com}) and Wind database (\url{www.wind.com.cn}), respectively.

\subsection{Application 1: International Portfolio Selection}\label{sec_5_1}

Our first application studies the international portfolio selection. As elaborated in Example \ref{exa1}, we consider a tensor time series $\mathcal{X}_t\in\mathbb{R}^{10 \times 11 \times 4}$ spanning from 2018/05/03 to 2022/12/03 with $1134$ observations in total, where the $(i_1, i_2, i_3)$th entry of $\mathcal{X}_t$ is the daily simple return (in percentage) of a style that consists of a stock from market $i_1$ in sector $i_2$ with the market capitalization level $i_3$\footnote{Note that the market capitalization level of a stock is determined by the data on 2018/05/03, and the stock to form the style is the one having the largest market capitalization in market $i_1$, sector $i_2$, and at the market capitalization level $i_3$.}. Here, the ten international stock markets (w.r.t. mode 1 of ``Market'') are Australia, Canada, China, German, Hong Kong, India, Japan, Korea, UK, and USA; the eleven sectors (w.r.t. mode 2 of ``Sector'') are Basic Materials, Consumer Cyclical, Financial Services, Real Estate, Consumer Defensive, Healthcare, Utilities, Communication Services, Energy, Industrial, and Technology; and the four market capitalization levels (w.r.t. mode 3 of ``Size'') are Mega (above 100B), Large (10B--100B), Middle (2B--10B), and Small (below 2B).

\begin{figure}[!htp]
	\centering
	\includegraphics[height=7cm, width=13.5cm]{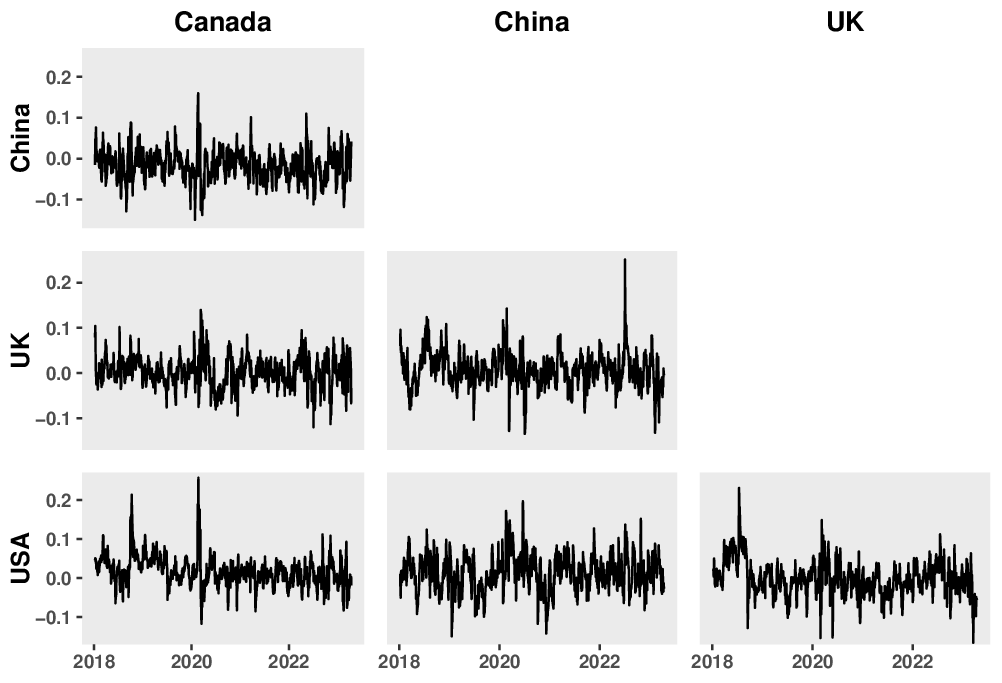}
        \includegraphics[height=7cm, width=13.5cm]{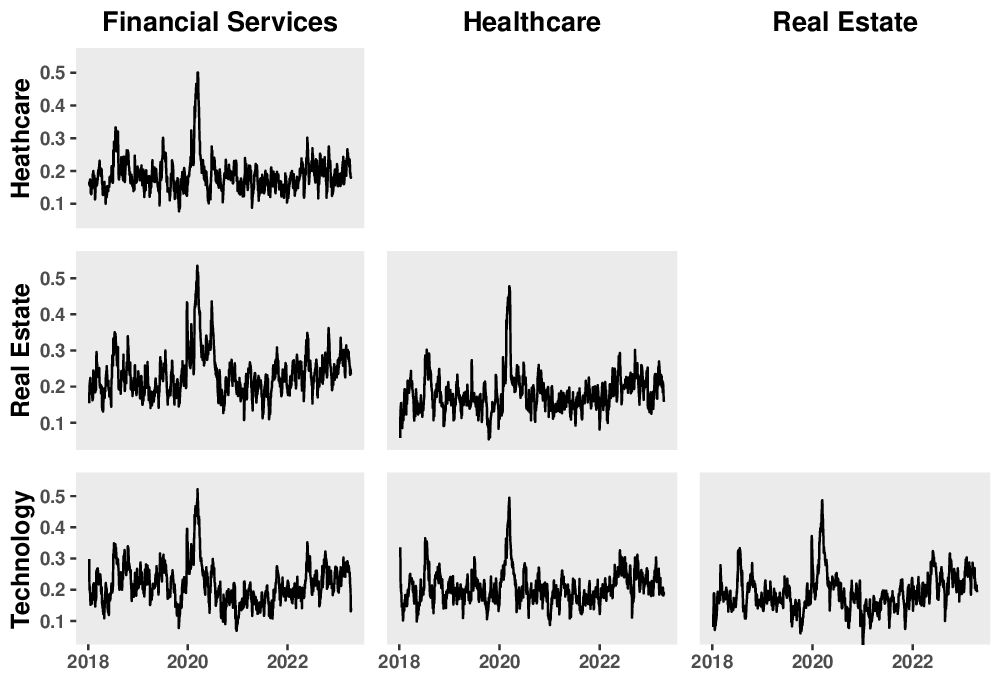}
        \includegraphics[height=7cm, width=13.5cm]{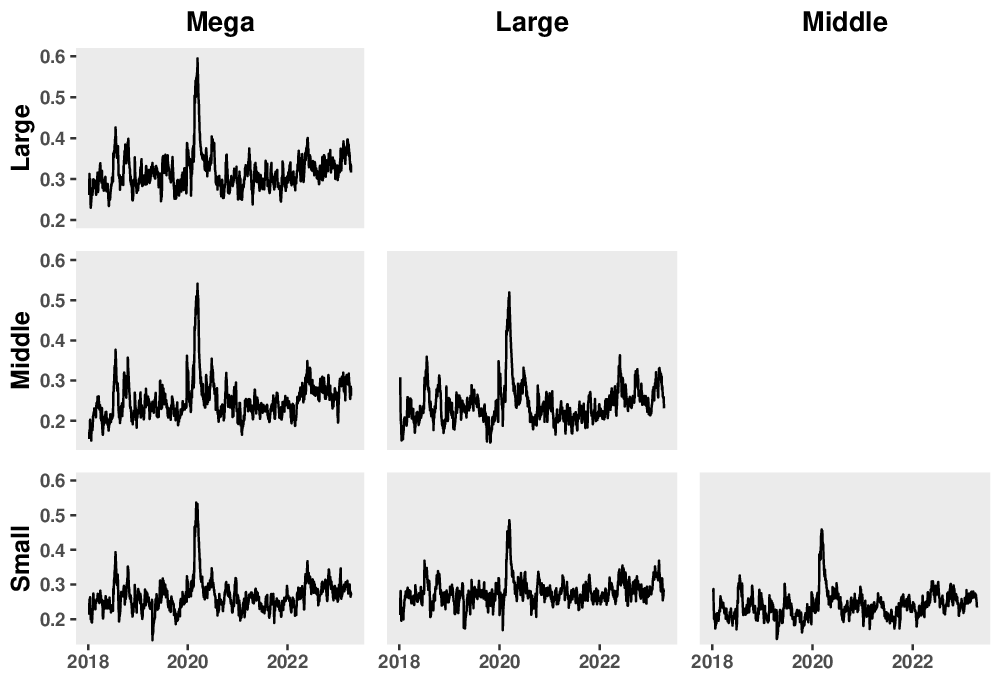}
	\caption{Part of estimated conditional correlation matrices: Top panel (market mode); Middle panel (sector mode); Bottom panel (size mode).}
	\label{Corr_Global_mode1}
\end{figure}

Since the sample autocorrelations of each entry series of $\mathcal{X}_t$ are insignificant, we apply the TDCC model to fit $\mathcal{X}_t$ (after removing a constant mean), where the intercept parameter matrix $\mathbf{C}_k$ is estimated by the sample covariance matrix method as in (\ref{new_est_C}). Based on the fitted TDCC model, we obtain the estimated mode-$k$ conditional correlation matrix $\mathbf{R}_{kt}$ for $k = 1,2,3$, and for ease of visualization, we only plot a part of $\mathbf{R}_{kt}$ in Figure \ref{Corr_Global_mode1}. From this figure, we find that the market-specific conditional correlations are generally very small (fluctuating around $0$), whereas the sector-specific and size-specific conditional correlations are relatively larger (but typically less than $0.3$). This finding shows that investing in various stock markets can be an effective way to diversify risk. Moreover, we detect that all series of conditional correlations in sector mode or size mode have a similar time trend, with a significant spike in February 2020 due to the systemic crisis triggered by COVID-19. It reveals that sector-specific and size-specific conditional correlations are good indicators for monitoring global financial risk.

\begin{figure}[!h]
 	\centering
 	\includegraphics[height=7cm, width=12cm]{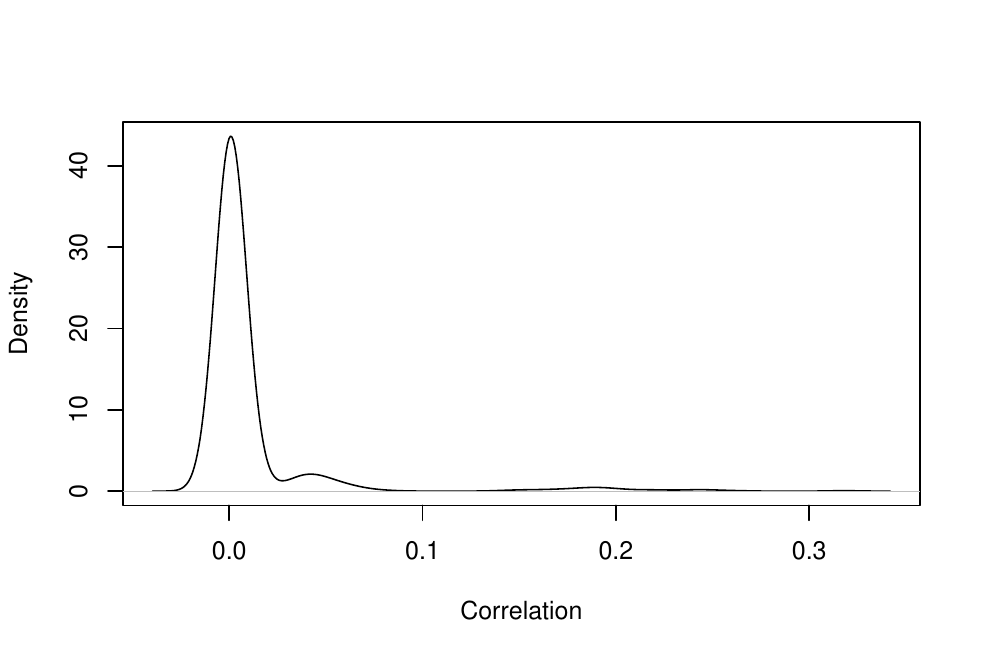}
 	\caption{Density of the 96,580 different time-averaged conditional correlations.}
 	\label{Density_Corr_Global}
 \end{figure}

Next, we investigate the conditional correlations among all 440 styles from the estimated conditional correlation matrix $\mathbf{R}_t$. Since there are 96,580 different series of conditional correlations in total, we visualize them collectively by plotting the density of their sample means in Figure \ref{Density_Corr_Global}. From this figure, it is evident that the averaged conditional correlations across all 440 styles are generally very small, supporting the use of these styles for proposing portfolios under the principle of ``Holy Grail of investing''.

Finally, we evaluate the performance of the TDCC model for the selection of international portfolios. Specifically, we split the entire time period into two parts: the training period with the first $T_{train}=630$ time points (around $2.5$ years), and the testing period with the remaining $T_{test}=504$ time points (around $2$ years). At each time point $t_0$ in the testing period, we fit the TDCC model based on the latest $T_{train}$ observations $\{\mathcal{X}_t\}_{t=t_0-T_{train}}^{t_0-1}$, and then use this fitted model to predict the conditional covariance matrix of $\textrm{vec}(\mathcal{X}_{t_0})$ by $\widehat{\mathbf{\Sigma}}_{t_0}$, where $\widehat{\mathbf{\Sigma}}_{t_0}$ is computed as in (\ref{sigma_exp}).

Using $\widehat{\mathbf{\Sigma}}_{t_0}$, we construct the unconstrained minimum-variance portfolio from $440$ styles by choosing the weight vector as
\begin{align}\label{c_MVP}
&\widehat{w}_{t_0}^{u}:=\argmin_{w_{t_0}'\mathbf{1}_{N} = 1} w_{t_0}'\widehat{\mathbf{\Sigma}}_{t_0}w_{t_0}=\frac{\widehat{\mathbf{\Sigma}}_{t_0}^{-1}\mathbf{1}_{N}}{\mathbf{1}_{N}' \widehat{\mathbf{\Sigma}}_{t_0}^{-1}\mathbf{1}_{N}}
\end{align}
for $N = 440$. If the short sales are not preferred or allowed, we can also select the constrained minimum-variance portfolio by choosing the weight vector as
\begin{align}\label{u_MVP}
&\widehat{w}_{t_0}^{c}:=\argmin_{w_{t_0}'\mathbf{1}_{N} = 1,\,\,w_{t_0}\geq \mathbf{0}} w_{t_0}'\widehat{\mathbf{\Sigma}}_{t_0}w_{t_0},
\end{align}
where $\widehat{w}_{t_0}^{c}$ has no closed form and needs to be computed by numerical optimization methods.
To evaluate the performance of the proposed unconstrained and constrained minimum-variance portfolios, we follow \cite{Engle2019Large} to
consider the out-of-sample averaged returns (AV), standard deviation of returns (SD), and information ratio (IR) defined by
\begin{align*}
\mbox{AV} = \frac{1}{T_{test}}\sum_{t_0}\widehat{w}_{t_0}'R_{t_0},\,\,\,
\mbox{SD} = \sqrt{\frac{1}{T_{test}-1}\sum_{t=1}^{T_{test}}(\widehat{w}_{t_0}'R_{t_0} - \mbox{AV})^2},\,\,\, \mbox{ and }\,\,\,
\mbox{IR}  = \frac{\mbox{AV}}{\mbox{SD}},
\end{align*}
respectively, where $\widehat{w}_{t_0}$ is $\widehat{w}_{t_0}^{u}$ (or $\widehat{w}_{t_0}^{c}$) with respect to unconstrained (or constrained) minimum-variance portfolio, and $R_{t_0}=\mathrm{vec}(\mathcal{X}_{t_0})$. As in Section \ref{sec_4_2_candidate}, we refer to the above approach as the TDCC-S method. Certainly, many other methods can also be used to predict $\mathbf{\Sigma}_{t_0}$, so the corresponding minimum-variance portfolios can be constructed similarly as in (\ref{c_MVP})--(\ref{u_MVP}). For comparison, we focus on the following eight additional methods: TDCC-LS, TDCC-NLS, MDCC1-NLS, MDCC2-NLS, MDCC3-NLS, VDCC-S, VDCC-LS, and VDCC-NLS (see Section \ref{sec_4_2_candidate} for their detailed elaborations).

Table \ref{Table_Portfolio1} presents the annualized average return (AV), standard deviation (SD), and information ratio (IR), where AV is multiplied by 252 and SD by $\sqrt{252}$, for both the unconstrained and constrained minimum-variance portfolios proposed by all nine methods. From this table, we find that for both unconstrained and constrained minimum-variance portfolios, the three TDCC-based methods have a very similar performance in terms of AV, SD, and IR, with the TDCC-NLS method showing a slight superiority over the TDCC-S and TDCC-LS methods. Moreover, we find that for the unconstrained (or constrained) minimum-variance portfolio, the TDCC-NLS method yields a 5.5\% (or 3.4\%) higher value of IR than the MDCC1-NLS method, which has the largest IR value within three MDCC-based methods. Compared with the VDCC-NLS method (i.e., the VDCC-based method with the largest IR value), the advantage of the TDCC-NLS method is even more substantial for the unconstrained minimum-variance portfolio, achieving a 78.1\% higher value of IR. The superior performance of the TDCC-NLS method over the MDCC1-NLS and VDCC-NLS methods indicates the need to maintain the tensor structure of the style return data for portfolio selection.

\begin{table}[htbp]
  \centering
  \small
  \caption{Minimum-variance portfolio performance based on international tensor return data}
  \setlength{\tabcolsep}{5.2mm}{
    \begin{tabular}{ccccccccc}
    \toprule
          &       & \multicolumn{3}{c}{Unconstrained} &       & \multicolumn{3}{c}{Constrained} \\
\cmidrule{3-5}\cmidrule{7-9}     \mbox{Method}     &       & AV    & SD    & IR    &       & AV    & SD    & IR \\
    \midrule
    TDCC-S &       & 6.75  & 3.97  & 1.70  &       & 7.16  & 3.95  & 1.81  \\
    TDCC-LS &       & 6.76 & 3.97  & 1.70  &       & 7.18  & 3.94 & 1.82  \\
    TDCC-NLS &       & 6.76 & 3.96 & 1.71 &       & 7.24  & 3.94 & 1.84 \\
    MDCC1-NLS &       & 6.68  & 4.12  & 1.62  &       & 7.29  & 4.09  & 1.78  \\
    MDCC2-NLS &       & 6.41  & 4.15  & 1.55  &       & 7.21  & 4.13  & 1.74  \\
    MDCC3-NLS &       & 4.81  & 4.28  & 1.12  &       & 6.88  & 4.15  & 1.66  \\
    VDCC-S &       & -3.48  & 7.62  & -0.46  &       & 5.66  & 4.43  & 1.28  \\
    VDCC-LS &       & 3.83  & 4.98  & 0.77  &       & 7.18  & 4.19  & 1.71  \\
    VDCC-NLS &       & 4.31  & 4.49  & 0.96  &       & 7.21  & 4.15  & 1.74  \\
    \bottomrule
    \end{tabular}%
    }
  \label{Table_Portfolio1}%
\end{table}%

\begin{figure}[!h]
    \centering
    \includegraphics[width=0.7\linewidth]{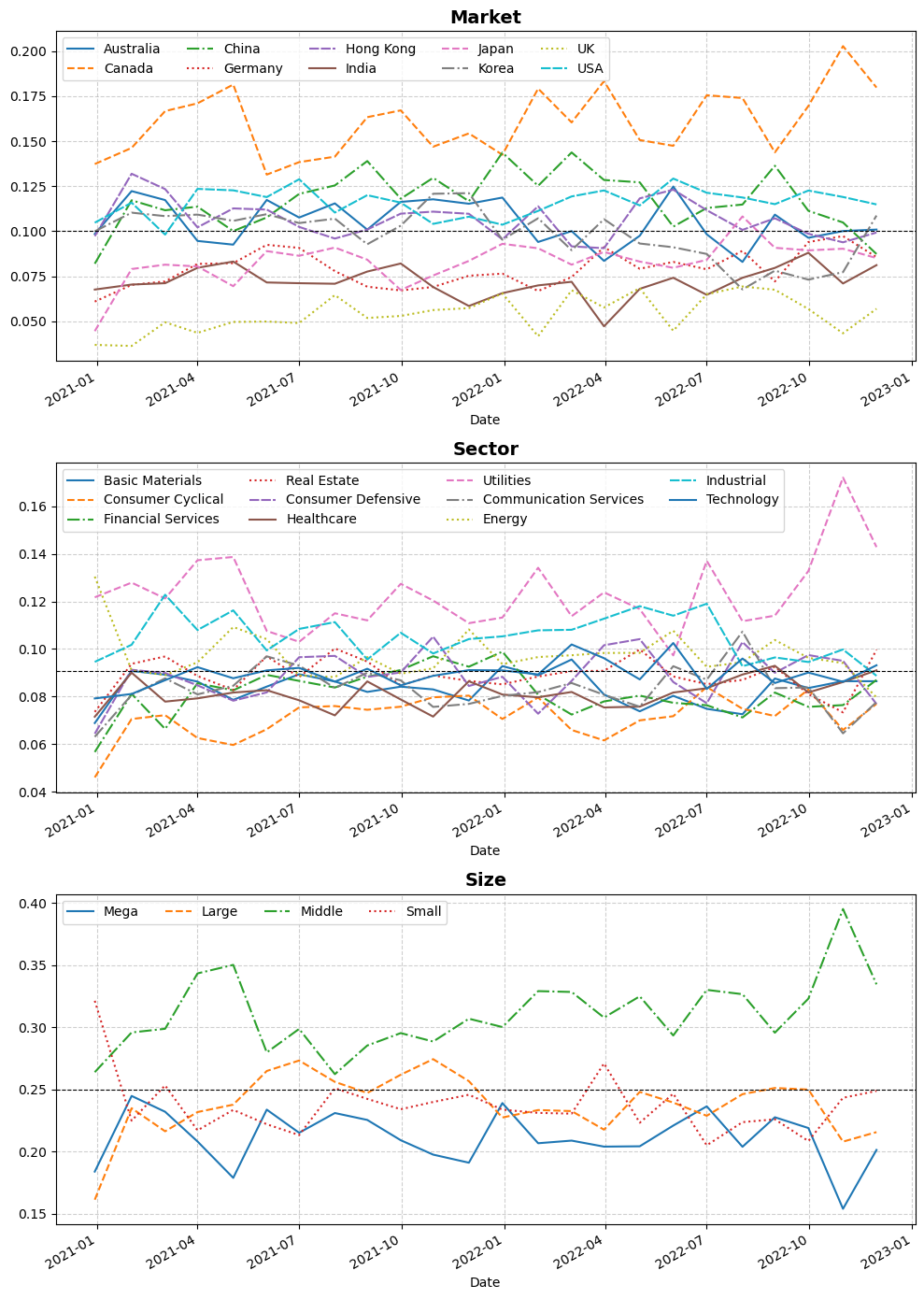}
    \caption{(Color online) The mode-specific averaged portfolio weights for the unconstrained minimum-variance portfolios selected by the TDCC-NLS method: Top panel (market mode); Middle panel (sector mode); Bottom panel (size mode). Here, the  horizontal dashed line represents the baseline equal weight for each mode.}
    \label{fig:app1_weight}
\end{figure}

In addition to portfolio performance, we are of interest to explore the evolution of style investment in our unconstrained minimum-variance portfolios selected by the TDCC-NLS method. To pursuit this goal, Figure \ref{fig:app1_weight} plots the mode-specific averaged portfolio weights, which exhibit the distribution of styles in each mode. From this figure, we observe that the distribution of styles remains stable overtime during the out-of-sample period. For the market mode, our minimum-variance portfolios keep investing the highest weight in Canada but the lowest weight in UK. This is probably because the Canadian market has negligible correlations with most of other considered markets, making it a good choice for diversifying risk, while the British market becomes less attractive due to the unstable economic and political conditions in the UK. For the sector mode, the Utilities and Industrial sectors receive higher weights than others. This is consistent with the risk-averse nature of these two sectors during the COVID-19 pandemic. For the size mode, the stocks with middle capitalization (mid-cap) are assigned with the highest weight overtime. Two explanations could be offered for this observation. First, we find from Figure \ref{Corr_Global_mode1} that mid-cap stocks exhibit relatively weak correlations with stocks in the mega and large capitalization categories, which, conversely, show strong correlations with each other. Therefore, mid-cap stocks are useful to diversify risk. Second, mid-cap stocks are typically favored by investors due to their growth potential, while being considered less risky than stocks with small capitalization.

\subsection{Application 2: Large Portfolio Selection}

\begin{figure}[!h]
    \centering
    \includegraphics[height=6cm, width=11cm]{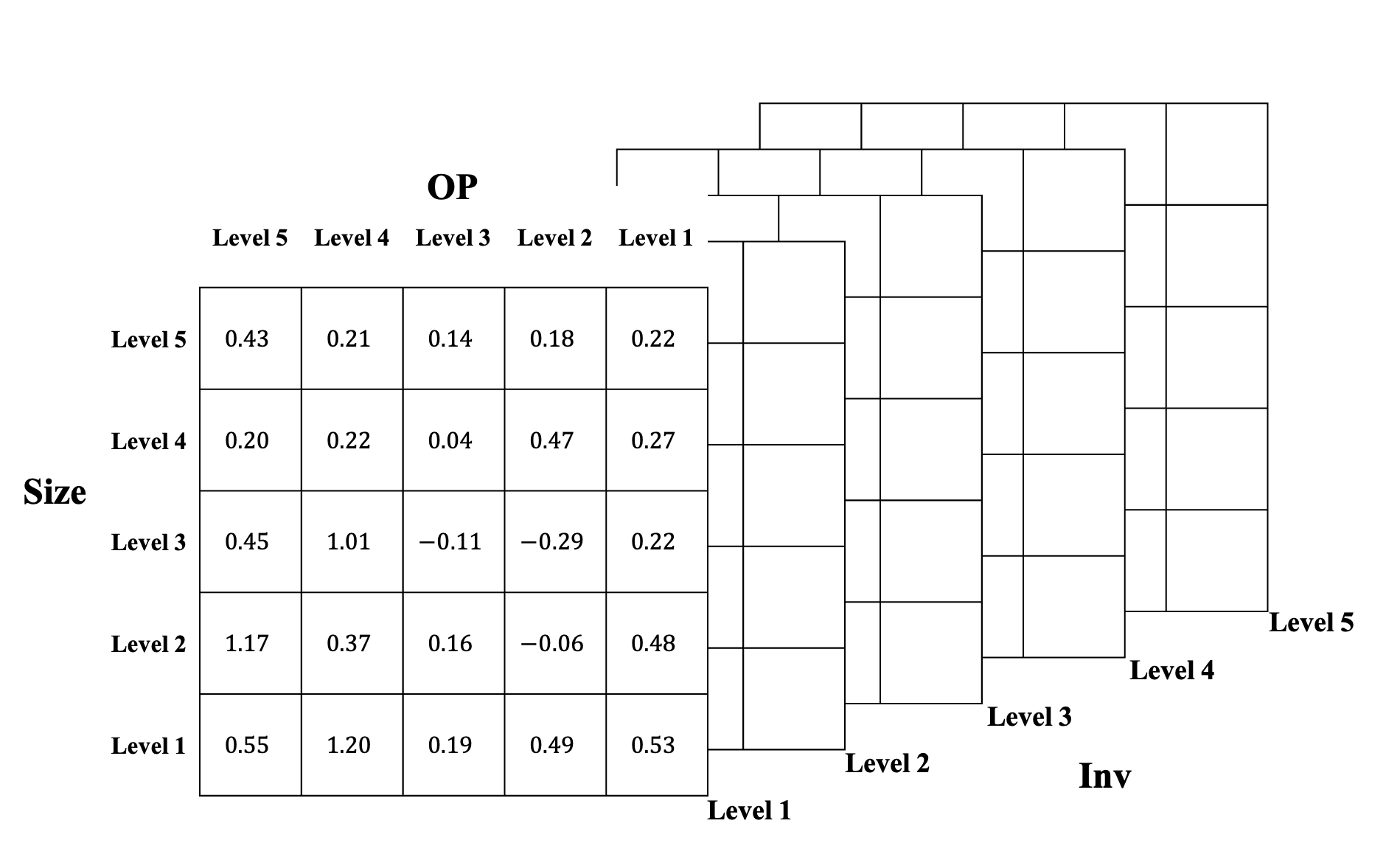}
    \caption{The first tensor observation of $\mathcal{X}_t$ on $2017/01/06$, consisting of 125 different weekly style returns (in percentage).}
    \label{Tensor_Factor}
\end{figure}

Our second application studies the large portfolio selection, based on $1800$ stocks with largest capitalization in Chinese stock market. Following Example \ref{exa2}, we construct a tensor time series $\mathcal{X}_t\in\mathbb{R}^{5 \times 5 \times 5}$ of weekly style returns from $2017/01/06$ to $2023/06/02$, with $328$ observations in total. Specifically, the $(i_1,i_2,i_3)$th entry of $\mathcal{X}_t$ is the weekly simple return (in percentage) of a style\footnote{Note that the weekly simple return of a style is computed as the average of weekly returns of all stocks in this style.} that contains all stocks having ``Size''= level $i_1$ in mode 1, ``OP''= level $i_2$ in mode 2, and ``Inv'' = level $i_3$ in mode 3. See a visualization of the first observation of $\mathcal{X}_t$ in Figure \ref{Tensor_Factor}.
Here, all of 125 styles are formed using the three characteristics ``Size'', ``OP'', and ``Inv'', each of which has five different levels (or states). To define the level of ``Size'' for each stock, we sort all 1800 stocks into deciles according to their values of size (from lowest to highest). Then, stocks in decile $2i-1$ and decile $2i$ have the level $i$ of ``Size'', for $i=1,\ldots, 5$. Similarly, we can define the level of ``OP'' and ``Inv'' for each stock. Since $\mathcal{X}_t$ exhibits a certain dynamic structure in its conditional mean, we fit the TDCC model to $\{\mathcal{X}_t\}_{t=1}^{328}$ after removing the conditional mean of $\mathcal{X}_t$ by a first-order TAR model (\citealp{li2021TAR}), where the intercept parameter matrix $\mathbf{C}_k$ is estimated by the sample covariance matrix method as in (\ref{new_est_C}). Based on the fitted TDCC model, some insightful findings can be obtained from the estimated $\mathbf{R}_{kt}$ and $\mathbf{R}_{t}$, and the details are deferred into Appendix C for saving the space.

Next, we assess the performance of the TDCC model for large portfolio selection in the Chinese stock market. Similar to Section \ref{sec_5_1}, we split the entire time period into two parts: the training period with the first $T_{train}=200$ time points (around 48 months), and the testing period with the remaining $T_{test}=128$ time points (around $32$ months). At each time point $t_0$ in the testing period, we use a fitted TDCC model (after removing the conditional mean via a first-order TAR model) based on the latest $T_{train}$ observations $\{\mathcal{X}_t\}_{t=t_0-T_{train}}^{t_0-1}$, to predict the conditional covariance matrix of $\textrm{vec}(\mathcal{X}_{t_0})$ by $\widehat{\mathbf{\Sigma}}_{t_0}$. As we aim to construct mean-variance portfolios, we also apply the fitted first-order TAR model to predict the conditional mean of $\textrm{vec}(\mathcal{X}_{t_0})$ by $\widehat{\mu}_{t_0}$.

Using $\widehat{\mu}_{t_0}$ and $\widehat{\mathbf{\Sigma}}_{t_0}$, we propose the unconstrained mean-variance portfolio from 125 styles by choosing the weight vector as
\begin{align}\label{c_Mean-Var}
&\widehat{w}_{t_0}^{u}:=\argmax_{w_{t_0}'\mathbf{1}_{N} = 1} \frac{w_{t_0}' \widehat{\mu}_{t_0}}{ w_{t_0}'\widehat{\mathbf{\Sigma}}_{t_0}w_{t_0} }
\end{align}
for $N = 125$. In addition, we can also propose the constrained mean-variance portfolio by choosing the weight vector as
\begin{align}\label{u_Mean-Var}
&\widehat{w}_{t_0}^{c}:=\argmax_{w_{t_0}'\mathbf{1}_{N} = 1,\,\,w_{t_0}\geq \mathbf{0}} \frac{w_{t_0}' \widehat{\mu}_{t_0}}{ w_{t_0}'\widehat{\mathbf{\Sigma}}_{t_0}w_{t_0} },
\end{align}
where $\widehat{w}_{t_0}^{c}$ has no closed form and needs to be computed by numerical optimization methods. Once the weight for each style is assigned as in (\ref{c_Mean-Var}) or (\ref{u_Mean-Var}), the assets in the style are invested in an equally-weighted way. As in Section \ref{sec_5_1}, the above approach is called the TDCC-S method, which is compared with eight different methods.

Table \ref{Table_Portfolio2} reports the annualized AV, SD, and IR of the unconstrained and constrained mean-variance portfolios proposed by all nine methods. The findings from Table \ref{Table_Portfolio2} are similar to those from Table \ref{Table_Portfolio1}, with some exceptions. First, TDCC-based unconstrained portfolios outperform MDCC-based and VDCC-based ones by a wide margin regarding the maximized value of IR. For example, the value of IR from the TDCC-NLS method is 22.5\% and 336.2\% higher than that from the MDCC1-NLC and VDCC-NLS methods, respectively. In terms of the constrained portfolios, the advantage of TCC-based methods over MDCC-based methods becomes weaker, while it still remains substantial compared with VDCC-based methods. Second, the three VDCC-based methods exhibit a very similar performance, suggesting that neither of the linear and nonlinear shrinkage methods learns $\mathbf{\Sigma}_t$ better than the sample covariance matrix method in this case.

\begin{table}[htbp]
  \centering
  \small
  \caption{Mean-variance portfolio performance based on Chinese tensor return data}
  \setlength{\tabcolsep}{5.2mm}{
    \begin{tabular}{ccccccccc}
    \toprule
          &       & \multicolumn{3}{c}{Unconstrained} &       & \multicolumn{3}{c}{Constrained} \\
\cmidrule{3-5}\cmidrule{7-9}    \mbox{Method}      &       & AV    & SD    & IR    &       & AV    & SD    & IR \\
    \midrule
     TDCC-S &       & 44.72 & 17.75 & 2.53 & &     19.42 & 12.06 &  1.62   \\
    TDCC-LS &      & 44.72 & 17.74 & 2.53 & &     19.42 & 12.05 &  1.62  \\
    TDCC-NLS &       & 44.72  & 17.74  & 2.53 &       & 19.42 & 12.05  & 1.62 \\
    MDCC1-NLS &       & 34.32  & 16.38  & 2.10  &       & 18.95  & 12.26  & 1.56  \\
    MDCC2-NLS &       & 37.32  & 18.91  & 1.97  &       & 18.52  & 12.17  & 1.54  \\
    MDCC3-NLS &       & 38.29  & 19.62  & 1.95  &       & 17.36  & 12.18  & 1.44  \\
    VDCC-S &       & 22.50  & 38.92  & 0.58  &       & 14.45  & 11.17  & 1.30  \\
    VDCC-LS &       & 23.10  & 39.72  & 0.58  &       & 14.28  & 10.93  & 1.31  \\
    VDCC-NLS &       & 24.52  & 42.03  & 0.58  &       & 14.38  & 10.83 & 1.32  \\
    \bottomrule
    \end{tabular}%
    }
  \label{Table_Portfolio2}%
\end{table}%

\begin{figure}[!h]
    \centering
    \includegraphics[width=0.7\linewidth]{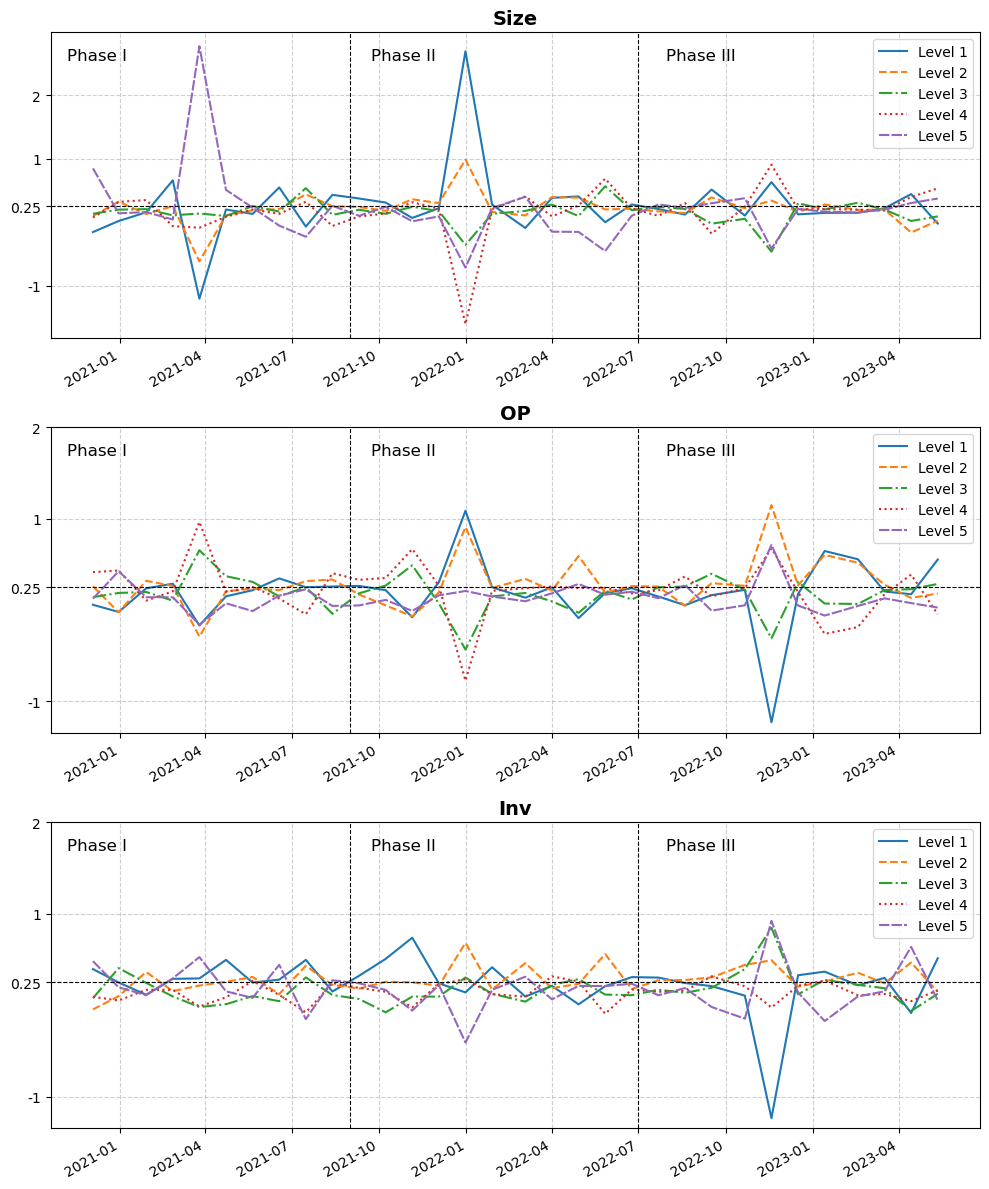}
    \caption{(Colar online) The mode-specific averaged portfolio weights for the unconstrained mean-variance portfolios selected by the TDCC-NLS method: Top panel (size mode); Middle panel (op mode); Bottom panel (inv mode). Here, the dashed horizontal line represents the baseline equal weight for each mode, and two dashed vertical lines divide the date axis into three phases.}
    \label{fig:size_weight}
\end{figure}

Finally, we pay attention to the evolution of style investment in our unconstrained mean-variance portfolios selected by the TDCC-NLS method. Figure~\ref{fig:size_weight} plots the mode-specific averaged portfolio weights, where, for better elaboration, its date axis is divided into three distinct phases: (I) December 2020 to September 2021, (II) October 2021 to June 2022, and (III) July 2022 to June 2023. In Phase I, we see a clear style shift in size mode starting from the beginning of 2021, when the Chinese government releases its first-quarter financial statements in March. As a feedback from this domestic shock, our TDCC-based portfolios favor (or disfavor) more styles with high (or small) value in size. In op and inv modes, there is no such apparent style shift. As the impact of this domestic shock becomes weak, it is expected to see that this style shift in size mode gradually diminishes.

In Phase II, we observe that in the first quarter of 2022, our TDCC-based portfolios change the distribution of styles to invest the most in stocks with small size, weak operating profitability, and conservative investment. This style shift in three modes is likely due to a very restrictive COVID-19 lockdown policy implemented by the Chinese government, causing large-scale manufacturing companies to shut down frequently and thus repel investors. In addition to the pandemic shock, another global event contributing to this shift is the outbreak of the Russia-Ukraine war in February 2022,
which makes prices of energy commodity surge.
Subsequently, the Federal Reserve raises interest rates by 25 basis points in March 2022. Both of these events heighten global inflation expectations and increase risks of stagflation, so stocks with low valuations and conservative investment profiles become more appealing to investors. In Phase III, we find the evident style shift in op and inv modes from the last quarter of 2022. Owing to the abolition of COVID-19 lockdown policy, this shift in two modes seems rational by investing growth styles with robust operating profitability and aggressive investment.

\section{Conclusions}\label{sec.concluding}

In this paper, we propose a new TDCC model to study the multi-way conditional heteroskedasticity of the tensor-valued time series of order $K$. The TDCC model step-wisely learns the conditional variances and mode-specific conditional correlation matrices. This provides us three practical advantages: First, it allows for the heterogeneity in the conditional variance of tensor entry; Second, it leads to informative interpretations regarding the mode-specific conditional correlations. Third, it facilitates a two-step estimation procedure with a manageable computational burden. When $K=1$, the TDCC model nests the widely used VDCC model (\citealp{engle2002dynamic}) as a special case. However, when $K\geq 2$, the TDCC model is more than a trivial generalization of the VDCC model, due to the use of two novel normalizations. Specifically, the TDCC model needs a trace-normalization to solve the issue of model identification, and at the same time, it must account for a dimension-normalization to specify the dynamics of conditional correlation matrix in each mode. The noteworthy endeavors we made for the case of $K\geq 2$ convey an important message: Generalizing a nonlinear vector model to its nonlinear tensor counterpart needs attentions.

Empirically, the TDCC model is motivated by the prevalent style investing, which creates a tensor of style returns with low correlations. By using the TDCC model to learn the conditional covariance matrix of all style returns, we obtain a new way to propose portfolios under the principle of ``Holy Grail of investing''. Our two applications, in terms of international portfolio selection and large portfolio selection, demonstrate that the TDCC-based method can outperform a range of competing methods, so they illustrate the importance of maintaining the tensor structure of style returns for learning conditional covariance matrix.

In future, many promising studies are anticipated.
First, as the VDCC model, exploring stationarity and establishing the asymptotic theory for the two-step estimators in the TDCC model remain open questions.
Second, one could extend the TDCC model with certain specific correlation specifications as in \cite{hafner2009generalized}, \cite{engle2012dynamic}, and \cite{hafner2022dynamic}. Third, a practical but challenging work is to construct styles through some machine learning methods, thereby possibly enhancing the effectiveness of TDCC-based portfolios.

\bibliographystyle{elsarticle-harv}
\bibliography{ref}

\end{document}